\documentclass[fleqn,usenatbib]{mnras}

\usepackage[T1]{fontenc}

\DeclareRobustCommand{\VAN}[3]{#2}
\let\VANthebibliography\thebibliography
\def\thebibliography{\DeclareRobustCommand{\VAN}[3]{##3}\VANthebibliography}

\usepackage{graphicx}	
\usepackage{amsmath}	
\usepackage{amssymb}	
\usepackage{txfonts}
\usepackage{enumerate}
\usepackage{microtype}
\usepackage[normalem]{ulem}
\usepackage[dvipsnames]{xcolor}

\usepackage{siunitx}
\usepackage{listings}

\DeclareSIUnit \h {\ensuremath{\mathit{h}}}
\DeclareSIUnit \parsec {pc}
\DeclareSIUnit \msol {\ensuremath{M_{\odot}}}

\newcommand{\rvtext}[1]{{#1}}

\newcommand{\myvec}[1]{\boldsymbol{#1}}

\newcommand{\subfind}{\textsc{subfind}}

\newcommand{\Mhm}{M_{\rm{hm}}}
\newcommand{\khm}{k_{\rm{hm}}}

\title[Mass functions for non-cold dark matter models]{Simulating the complexity of the dark matter sheet II: halo and subhalo mass functions for non-cold dark matter models}

\author[J. St\"ucker et al.]{
Jens St\"ucker,$^{1}$\thanks{E-mail: jstuecker@dipc.org}
Raul E. Angulo$^{1,2}$
Oliver Hahn,$^{3,4}$
and Simon D.M. White$^{5}$
\\
$^{1}$Donostia International Physics Centre (DIPC), Paseo Manuel de Lardizabal 4, 20018 Donostia-San Sebastian, Spain.\\
$^{2}$IKERBASQUE, Basque Foundation for Science, E-48013, Bilbao, Spain. \\
$^{3}$Department of Astrophysics, University of Vienna, T\"urkenschanzstrasse 17, 1180 Vienna, Austria \\
$^{4}$Department of Mathematics, University of Vienna, Oskar-Morgenstern-Platz 1, 1090 Vienna, Austria \\
$^{5}$Max Planck Institute for Astrophysics, Karl-Schwarzschild-Str. 1, 85741 Garching, Germany.
}

\date{Accepted XXX. Received YYY; in original form ZZZ}

\pubyear{2021}

\begin{document}
\label{firstpage}
\pagerange{\pageref{firstpage}--\pageref{lastpage}}
\maketitle

\begin{abstract}
We present ``sheet+release'' simulations that reliably follow the evolution of dark matter structure at and below the dark matter free-streaming scale, where
instabilities in traditional N-body simulations create a large population of spurious artificial haloes. Our simulations sample a large range of power-spectrum cutoff functions, parameterized through the half-mode scale $k_{\rm{hm}}$ and a slope parameter $\beta$. This parameter space can represent many non-cold dark matter models, including thermal relic warm dark matter, sterile-neutrinos, fuzzy dark matter, and a significant fraction of ETHOS models. Combining these simulations with additional N-body simulations, we find the following results. (1) Even after eliminating spurious haloes, the halo mass function in the strongly suppressed regime ($n_{\rm{X}}/n_{\rm{CDM}} < 5\%$) remains uncertain because it depends strongly on the definition of a halo. At these mass scales traditional halo finders primarily identify overdensities that are unbound, highly elongated, dominated by tidal fields, or far from virialized.  (2) The regime where the suppression is smaller than a factor of 20 is quite robust to these uncertainties, however, and can be inferred reliably from suitable N-body simulations. (3)  Parameterizing the suppression in the halo- and subhalo mass functions through the scales where the suppression reaches $20\%$, $50\%$ and $80\%$, we provide simple formulae which enable predictions for many non-cold dark matter models. (4) The halo mass-concentration relations in our sheet+release simulations agree well with previous results based on N-body simulations. (5) In general, we confirm the validity of previous N-body studies of warm dark matter models, largely eliminating concerns about the effects of artificial haloes.
\end{abstract}

\begin{keywords}
methods: numerical -- cosmology: large-scale structure of Universe  --  dark matter
\end{keywords}



\section{Introduction}
While the cosmological parameters of our Universe have been measured to an astonishing accuracy, some key questions about our Universe remain unanswered. One of those key questions is: \emph{What is the nature of Dark Matter?} So far, dark matter seems observationally consistent with a perfectly cold, collisionless fluid. However, most particle dark matter candidates are expected to deviate from this fiducial scenario at some scale. Such deviations can for example be caused by the \emph{warmth}\footnote{more accurately: the primordial velocity dispersion} of dark matter, its \emph{self-interaction} or its \emph{quantum nature}. 

Dark matter cannot only be investigated by possible interactions with known species in a particle physics lab, but also by the imprints that its properties leave on the formation of structures. Different properties will in general cause different behaviour for the detailed small-scale physics. However, at first order, the main effect of these properties onto the formation of structures is the suppression of the power spectrum on small scales \citep{murgia2017}. The usually considered canonical case for a dark matter candidate with such a small-scale suppression is thermal relic warm dark matter. However, for example, also fuzzy dark matter \citep[e.g.][]{Niemeyer_2020} and many ETHOS models \citep{Racine_2016} exhibit a small-scale suppression, just with a different precise shape than canonical warm dark matter. This suppression will already go a long way for explaining a lot of the difference to cold dark matter scenarios -- for example in the halo mass function. Here, we will  refer to any dark matter model with a small-scale cutoff as ``non-cold'' dark matter, and we aim to understand the general impact of a small-scale suppression in the power spectrum.

There exist several promising probes that can be used to constrain (or possibly detect) non-cold dark matter candidates observationally. These include the absorption pattern in the Lyman-$\alpha$ forest \citep{viel_2013, murgia_2018}, perturbations in arcs of strongly lensed galaxies \citep{vegetti_2014, vegetti_2018}, flux-ratio anomalies of quadruply lensed quasars \citep{gilman_2020}, number counts of Milky-Way satellites \citep{lovell2014, newton_2020} and gaps in tidal streams from satellite galaxies \citep{Yoon_2011, banik_2018, banik_2019}. In all of these cases the aim is to distinguish cold from non-cold dark matter by either detecting small structures, as expected in cold dark matter scenarios, or by failing to detect them in expected numbers, if the primordial power-spectrum is suppressed on small scales relative to CDM  \citep[see e.g.][for an overview]{enzi_2020}.

All of these observational constraints depend crucially on the ability to predict how the properties and the abundance of small-scale structure depend on the dark matter candidate. For example, the modelling of flux-ratio anomalies simultaneously requires models for the halo mass function, for the subhalo mass function and for the concentration-mass relation in non-cold dark matter cosmologies \citep{gilman_2020} and may even require models for non-halo structures \citep{richardson_2021}. While it is commonly assumed that N-body simulations can produce converged predictions for these quantities, this has never been proven reliably. \citet{wang2007} found  that N-body simulations of warm dark matter universes fragment even well below the free-streaming scale into large numbers of artifical haloes. While such artificial haloes can be at least partially removed in \rvtext{post-processing \citep{lovell2014, Agarwal_2015}}, it is not clear whether this approach leads to the same result as would have been found in the absence of the artifact. For example, the small artificial haloes may alter the accretion history of larger haloes and ultimately modify their concentrations. Since virtually all quantitative constraints on the warmth of dark matter depend crucially on the predictions from N-body simulations, it is important to test N-body
results using independent methods.

Recently, such an alternative approach to simulating collisionless cosmological evolution has been developed, so-called  ``sheet-based'' simulation methods \citep{abel2012, shandarin2012, hahn2013, hahn2016, sousbie2016, stuecker2020complexity}. In such simulations an accurate density estimate without discreteness noise is obtained by reconstructing the dark matter sheet in phase space. This allows simulations of cosmologies with a small-scale cutoff in the power spectrum without any artificial fragmentation \citep{angulo2013}. However, it comes at the cost of intractably complex dynamics inside of haloes -- leading either to biased density estimates at their centers  \citep{angulo2013} or to simulations that cannot be extended until $a=1$ because of exploding refinement costs  \citep{sousbie2016}. In \citet{stuecker2020complexity} we presented a solution to this dilemma; our `sheet + release' (S+R) scheme combines the benefits of the N-body and sheet approaches, by using sheet reconstructions wherever possible, but switching to an N-body representation where the dynamics becomes too complex (in the inner regions of haloes).

Here, we will present the first large set of simulations carried out with the S+R method. These simulations span a wide parameter space of dark matter models with a primordial power spectrum cutoff. Our goal here is two-fold. First, we aim to understand the physical implications of the cutoff. Of course, these effects have been investigated in previous studies \citep[e.g.][]{angulo2013, lovell2014, ludlow2016}, but never with a scheme that is reliable from scales at and below the cutoff up to the dense centers of haloes. Here we can present the first analysis of simulations that are free of numerical artifacts. We use this to test the reliability of previous results based on N-body simulations, and we can point out the conditions under which  the application of N-body results is problematic. We note that \citet{colombi_2021} has already addressed a similar question by comparing \textsc{ColDICE} \citep{sousbie2016} sheet-based simulations with N-body simulations that use a large softening length (of at least one mean particle separation) in a detailed phase space analysis. However, here we are instead interested whether classical N-body simulations with a small softening length -- as employed in the majority of previous investigations -- can give robust results on the level of coarse-grained summary statistics, such as the halo mass function and the mass concentration relation.

Further, we develop comprehensive quantitative descriptions of the halo mass function (HMF) and the subhalo mass function (SHMF) in cosmologies with a cutoff. Inspired by the study of \citet{murgia2017}, we capture the different cutoffs that power spectra can exhibit due to the particle nature of dark matter in a small set of parameters. Here, we use two, one indicating the scale of the cutoff and the other its sharpness. From our simulations we infer a simple functional form for the suppression in the (S)HMF as a function of our two parameters. This description can be used to evaluate the (S)HMF for almost any kind of dark matter model which exhibits small-scale suppression. We hope that these results will facilitate observational studies aimed at inferring constraints on warm dark matter as well as on the parameter space of other non-cold dark matter models.

This article is structured as follows: In Section \ref{sec:simulations} we present the simulations that we have run for this article and show how most non-cold dark matter cutoffs can be captured in a two-parameter description. In Section \ref{sec:uncertaintyofMF} we investigate the halo mass function on an absolute scale and quantify the uncertainties that arise due to issues about which objects should be considered halos. Further, we compare the results of our S+R scheme with those from N-body simulations. In Section \ref{sec:quantitativencdm} we quantify the relative suppression of the (S)HMF and provide simple descriptions that can be rescaled to many different dark matter models. Finally, in Section \ref{sec:massconcentration} we have a brief look at the mass concentration relations in our S+R simulations, comparing them to the relations that have previously been inferred from N-body simulations.

\begin{figure*}
    \centering
    \includegraphics[width=\textwidth]{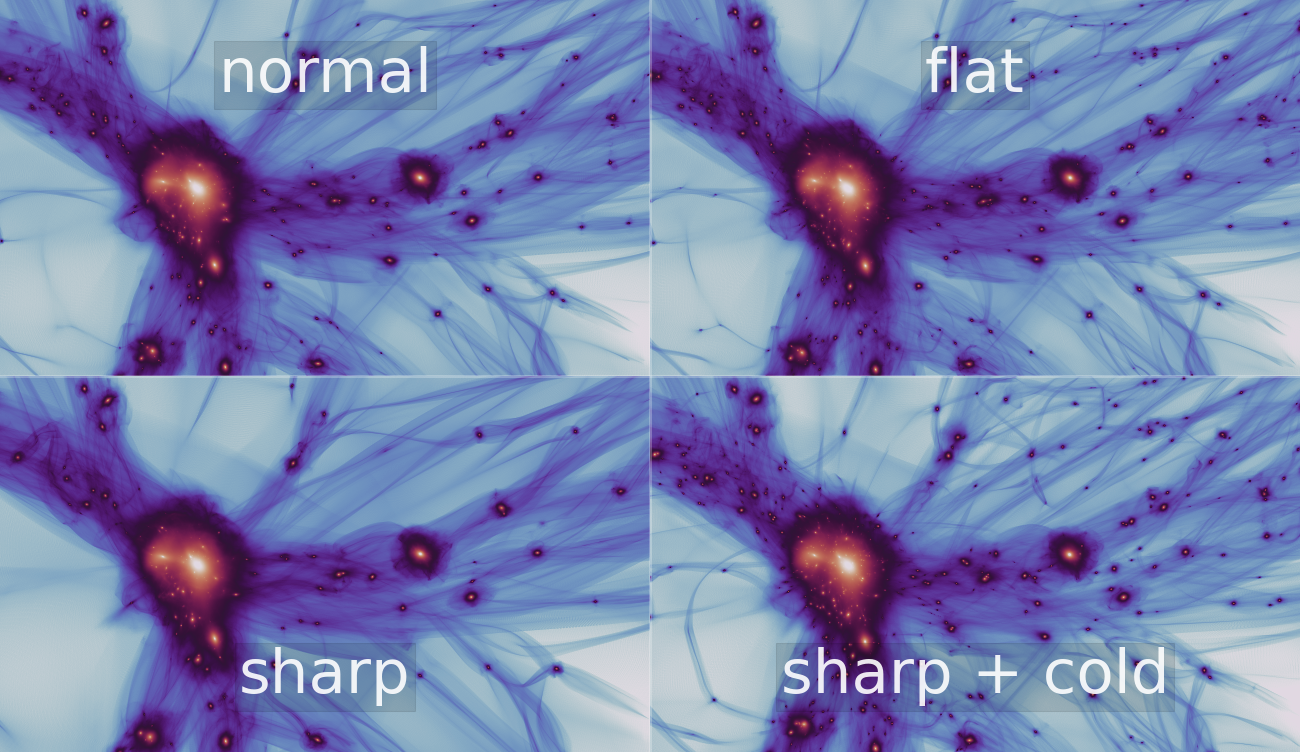}\\
    \includegraphics[width=\textwidth]{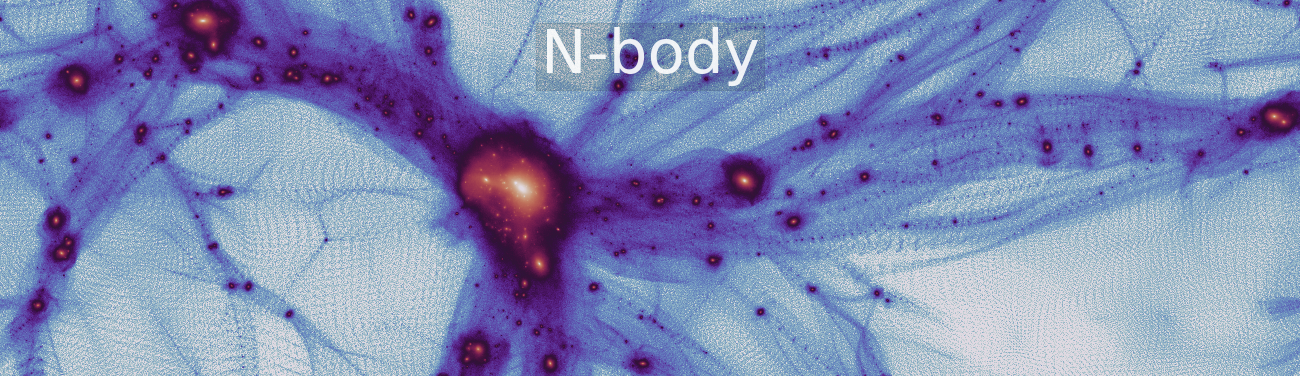}
    \caption{Visualization of a region of the different sheet + release simulations presented in this article (top four panels) and of an N-body simulation ffor the `normal' cutoff case (the bottom panel). Three panels show cases with differing slope $\beta$ but the same scale for the cutoff, while one shows results for the `normal' slope but a smaller cutoff in mass (larger $\khm$). The sheet + release simulations produce a high-quality density field with no sign of the artificial small-scale structures which are quite prominent in the N-body case. A video (showing the 1.75kev and 1kev cases with $\beta=6$) can be found here: \href{https://bacco.dipc.org/ncdm.html}{https://bacco.dipc.org/ncdm.html}}
    \label{fig:simulation_image}
\end{figure*}

\section{Simulations} \label{sec:simulations}

We set up a suite of 'sheet + release' (S+R) simulations which test the effect of various power-spectrum cutoffs of the kind obtained generically in non-cold dark matter models. The goal is to use our novel S+R simulation scheme to create reliable fragmentation-free simulations of such universes and to measure how the \rvtext{halo- and subhalo}-mass functions depend on the parameters that define the power spectrum cutoff. We will explain our parameterization of the space of initial power spectra in  Section \ref{sec:transferfunctions} and in Section \ref{sec:sheetplusrelease} we review the main points of the S+R simulation scheme presented previously by \citet{stuecker2020complexity}. 

We show visualizations of part of our simulations in Figure \ref{fig:simulation_image}. Clearly our S+R technique results in a smoother and better defined density field which does not suffer from artificial fragmentation -- unlike the corresponding N-body simulation. Note, however, that in high-density regions the simulations are indistinguishable in these images

\subsection{Transfer Functions} \label{sec:transferfunctions}
\begin{figure}
    \centering
    \includegraphics[width=\columnwidth]{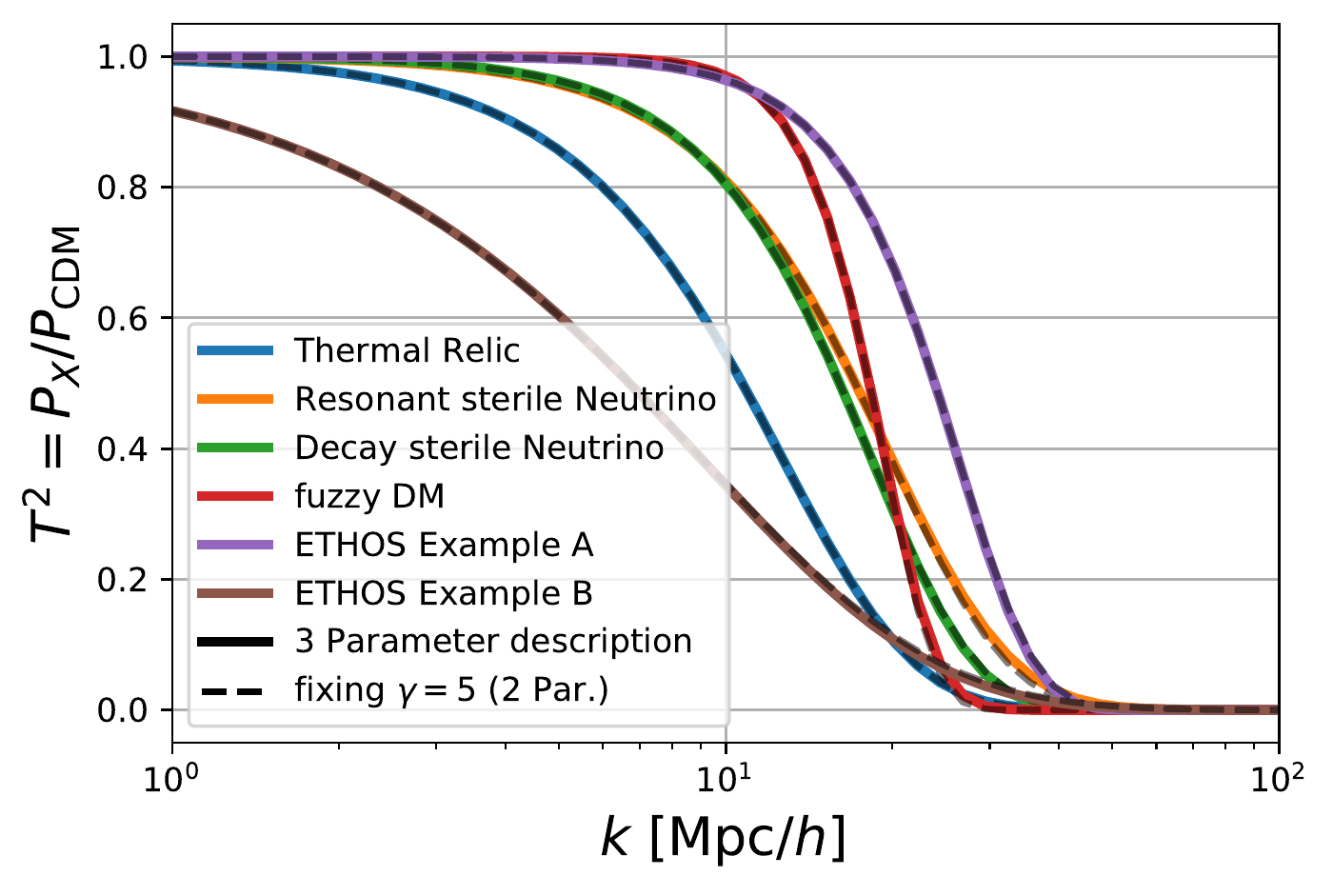}
    \caption{Examples of different dark matter models that can be described through our 2-dimensional parameter space $(\beta, \khm)$. See Figure \ref{fig:murgia_fits} for a more detailed version of this Figure that also displays the corresponding best-fit parameters. Almost all dark matter models with a small-scale cutoff are well approximated by a two-parameter fit of the cutoff scale $\khm$ and the steepness parameter $\beta$.}
    \label{fig:transfermodelssimple}
\end{figure}

Inspired by \citet{murgia2017}, we parameterize the ratio between a non-cold dark matter (NCDM) power spectrum $P_X(k)$ and the cold dark matter (CDM) power spectrum $P_{\rm CDM}(k)$ through the NCDM transfer function
\begin{align}
    T(k) &:= \sqrt{ \frac{P_{X}(k)}{P_{\rm{CDM}}(k)} } 
         \;\simeq\; \left(1 + \left(\alpha k\right)^\beta\right)^{-\gamma}, \label{eqn:transferabg}
\end{align}
where $k$ is the comoving wave number. We find that the parameters $\alpha$, $\beta$, $\gamma$ are typically highly degenerate. Therefore we re-express $\alpha$ through the half-mode $\khm$ which indicates the scale where the transfer-function $T$ is suppressed by a factor $2$:
\begin{align}
    \alpha &= \frac{\left(2^{\frac{1}{\gamma}} - 1 \right)^{\frac{1}{\beta}}}{\khm}
\end{align}
leaving us with the three parameters $(\khm, \beta, \gamma)$. However, we find that most NCDM models are well fit by a two parameter model fixing $\gamma=5$. We demonstrate this for a few examples in Figure \ref{fig:transfermodelssimple} where we fitted such a two parameter model to a large variety of different NCDM models. Note that the solid lines are the three-parameter models used in \citet{murgia2017}, and the dashed lines are our two parameter approximations. This means that most dark matter models with a small-scale cutoff -- like thermal relics, resonant sterile neutrinos, decay sterile neutrinos, fuzzy dark matter, and many ETHOS models \citep{Racine_2016} -- can be well described by the parameters $\khm$, controlling the cutoff scale, and $\beta$, controlling the steepness of the cutoff. For a more detailed and quantitative evaluation please consider Figure \ref{fig:murgia_fits} in the appendix.  Generally, we find that only mixed dark matter models and ETHOS models with a significant dark acoustic oscillation are not always well modeled by our two parameter functions.

\begin{table}
    \centering
    \begin{tabular}{c||r|c|c||c|c}
                       Label &      $\khm$ &       $\beta$ &      $\gamma$ &  $k_{\sqrt{1/2}}$ &    $\Mhm$\\
        \hline
            1keV sharp * &           5.95 &          6 &          5 &        5.3 &     5.2e+10\\
            1keV   * &         7.59 &          2 &          5 &        5.3 &     2.5e+10\\
            1keV flat * &          8.57 &        1.5 &          5 &        5.3 &     1.8e+10\\
            1.75keV sharp * &      11.33 &          6 &          5 &       10.0 &     7.6e+09\\
            \hline \hline
            1.75keV      &      14.44 &          2 &          5 &       10.0 &     3.7e+09\\
            1.75keV flat      &      16.31 &        1.5 &          5 &       10.0 &     2.5e+09\\
            3keV sharp     &      21.06 &          6 &          5 &       18.7 &     1.2e+09\\
            3keV     &      26.85 &          2 &          5 &       18.7 &     5.7e+08\\
            3keV flat     &      30.31 &        1.5 &          5 &       18.7 &     4.0e+08\\
            \hline
            1keV $\beta=4$    &       6.33 &          4 &          5 &        5.3 &      4.3e+10\\
            1keV $\beta=3$      &      6.72 &          3 &          5 &        5.3 &      3.6e+10\\
            1keV $\beta=2.5$    &       7.06 &        2.5 &          5 &        5.3 &      3.1e+10\\
            1keV $\beta=1.75$     &       7.99 &       1.75 &          5 &        5.3 &      2.2e+10\\
            
            1keV $\gamma=2$     &       7.80 &          2 &          2 &        5.3 &     2.3e+10\\
            1keV $\gamma=10$    &       7.52 &          2 &         10 &        5.3 &     2.6e+10\\
        \hline
           CDM &           - &          - &          - &        -&     -\\
    \end{tabular}
    \caption{Overview of the power spectra parameter of the simulations that are presented in this paper. The units of $\khm$ and $k_{\sqrt{1/2}}$ are  $h/\rm{Mpc}$ and $\Mhm$ is in units of $M_\odot / h$. Simulations marked by an asterisk have been run with the S+R (sheet +  release) and the N-body method whereas the remaining simulations have only been run with the N-body scheme. The S+R (sheet + release) simulations are used for most of the analysis and for modelling the mass functions. The N-body simulations are used for comparison and to validate the results for other sets of parameters.}
    \label{tab:sims_cutoffs}
\end{table}

Therefore, we set up a suite of simulations which explores cutoff functions with fixed $\gamma = 5$, but varying parameters $\khm$ and $\beta$. In this article we will mostly focus on four sheet + release simulations (S+R) that explore this space. However, we will also validate our results and extrapolations through a larger set of N-body simulations which include additional different values of $\khm$ and $\beta$ and further explore the difference found when using $\gamma = 2$ and $\gamma = 10$. See Table \ref{tab:sims_cutoffs} for a full overview over our set of simulations. Note that we label the simulations e.g. as 1 keV 'sharp'/'normal'/'flat', indicating that these simulations have a similar cutoff scale to a thermal relic with $m_X$ = 1 keV, but with a sharper or flatter cutoff with respect to the thermal relic case. This approximate correspondence is just pointed out to guide the intuition, but it is not relevant for the results presented in this article.

Following \citet{schneider2012} the half-mode mass can be defined through
\begin{align}
    \Mhm &:= \frac{4 \pi}{3} \rho_0 \left(\frac{\pi}{\khm} \right)^3 \label{eqn:halfmode}
\end{align}
 where $\rho_0$ is the mean matter density of the universe at $z=0$. We list $\Mhm$ also in Table \ref{tab:sims_cutoffs}. $\Mhm$ is formally defined through the power spectrum, but empirically one finds that it roughly corresponds to the mass-scale where the halo mass function is suppressed by about a factor 2 or 3 with respect to the CDM case. However, note that the definitions of the half-mode scale  and half-mode mass are somewhat arbitrary. For example, one could also have used the mode $k_{\sqrt{1/2}}$ where the power spectrum (and not the transfer function) is suppressed by a factor 2 -- see also \citet{murgia2017}. We also list this scale in Table \ref{tab:sims_cutoffs}, since this is the scale which we originally kept fixed across different simulations. However, throughout this article we will express our results in terms of the more common definitions of $\khm$ and $\Mhm$ to make the comparison with literature results easier.
 
 \begin{figure}
    \centering
    \includegraphics[width=\columnwidth]{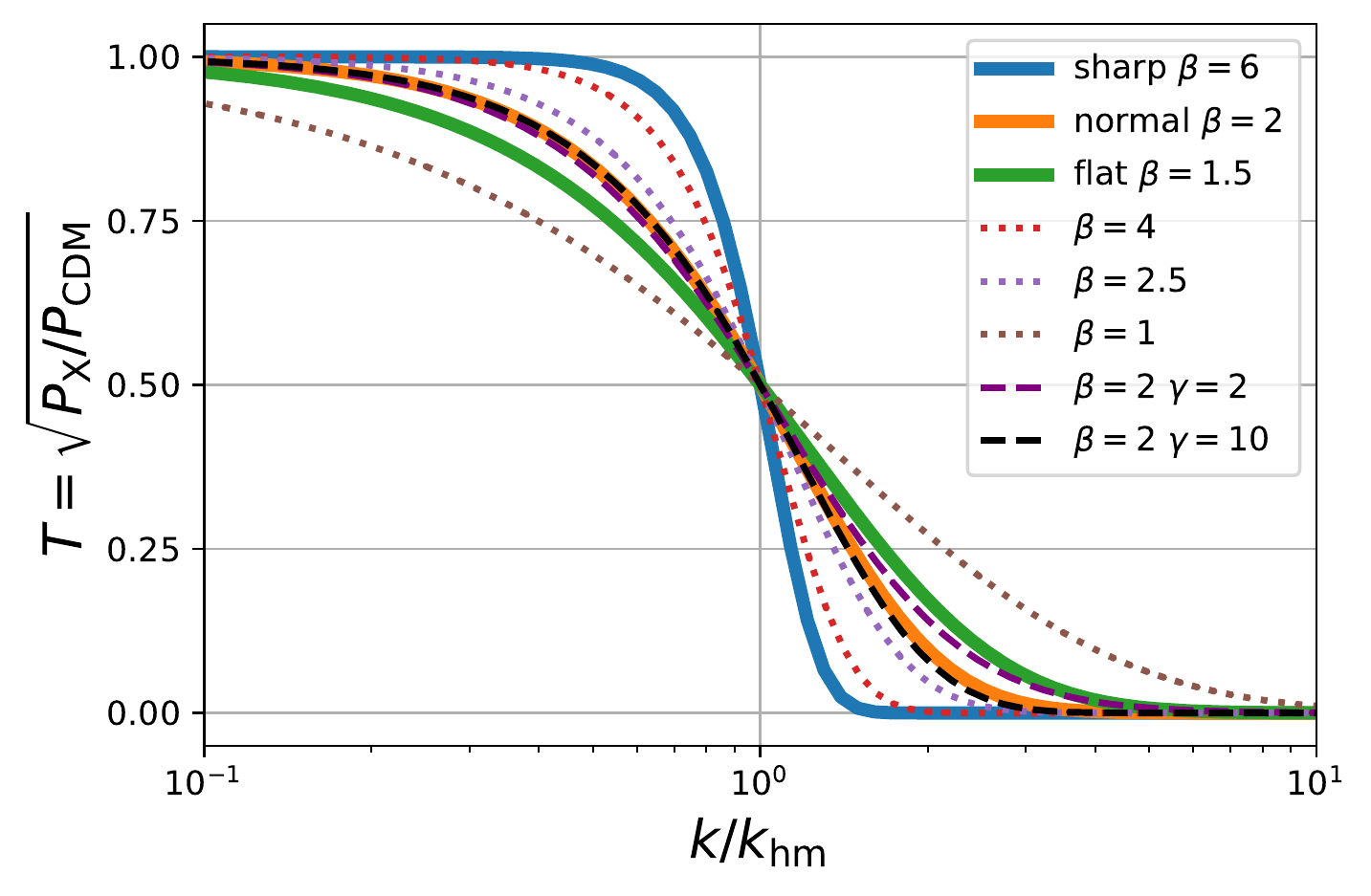}
    \caption{A subset of the transfer functions used in the simulations. The solid lines show the transfer-functions used by the S+R simulations. The dotted lines show simulations with other values $\beta$ which have been run as N-body simulations for validation purposes. The dashed lines vary $\gamma$. It seems clear that varying $\gamma$ at a fixed $\beta$ and $\khm$ induces only minor differences in the transfer-function. See also Table \ref{tab:sims_cutoffs}.}
    \label{fig:sims_transfer}
\end{figure}

In Figure \ref{fig:sims_transfer} we show the transfer functions of these simulations relative to $\khm$.

\subsection{The Sheet + Release simulation method} \label{sec:sheetplusrelease}
Here we briefly review the main aspects of the sheet + release (S+R) simulation scheme which was first introduced in \citet{stuecker2020complexity}. The main purpose of this scheme is to reliably simulate universes with a small-scale cutoff in the power-spectrum -- like warm dark matter universes or other non-cold dark matter scenarios.

As discussed in the introduction, N-body simulations of warm dark matter universes suffer from severe discreteness effects \citep{wang2007}. These discreteness effects are avoided by 'sheet' simulation schemes \citep{hahn2013, hahn2016, sousbie2016}, which use superior density estimates as can be seen in the top panels of Figure \ref{fig:simulation_image} in comparison to an N-body simulation in the bottom panel. However, pure 'sheet' simulations come at the cost of intractable dynamics inside of haloes. 

The S+R simulation scheme addresses these issues by combining the benefits of N-body and sheet simulation techniques \citep{stuecker2020complexity}. In the S+R simulation scheme, initially all mass is followed through a sheet interpolation approach. However, in regions where the dynamics becomes too complex -- that is in haloes -- mass elements are 'released', which means that their mass is subsequently followed through a set of N-body particles. Further, we have shown in \citet{stuecker2020complexity} that one can discretize the density field in this scheme well through an oct-tree of cubes where the density distribution in each cube can be well inferred through a joint assignment of sheet and N-body mass elements. 

The S+R is the first scheme that can reliably simulate warm dark matter universes -- all the way from large scale structures without any artificial haloes down to the very centers of haloes. In this paper we present the first set of simulations run with this scheme and can therefore test many of the predictions that have been previously made through N-body \rvtext{simulations \citep[e.g.][]{lovell2014, ludlow2016, Corasaniti_2017, Lovell2020, bohr2021}}  through an independent scheme.

\subsection{Simulation parameters}

For the simulations we choose the following parameters: The boxsize is fixed to $L = 20 \rm{Mpc} / h$ and the softening scale is set to $\epsilon = 0.5\rm{kpc}/h$ for both the N-body and the S+R simulations. For the N-body simulations we use $N = 512^3$ particles and for the sheet simulations, we use $N_T = 256^3$ sheet-tracing particles, which are used to reconstruct the interpolated sheet, and $N = 512^3$ 'silent' particles which are turned into N-body particles for released mass elements. Therefore the S+R simulations have 'infinite' mass resolution in regions where the sheet can be reconstructed and have the same mass resolution as the N-body simulations ($M_{\rm part} = 5.0 \cdot 10^{6} M_\odot / h$) in released regions. For the structure finding algorithms we only use the $512^3$ silent particles for the S+R simulations, so that the results can be directly compared with the N-body simulations.

Since the S+R simulations only sample the initial conditions grid at a resolution of $256^3$ they have a smaller Nyquist frequency ($k_{\rm{Ny}} = \SI{40.2}{\per \h \mega \parsec}$) than the reference N-body simulations. To make sure that the cutoff is well resolved we required that the total variance of the density field when only considering the power spectrum up to half of the Nyquist frequency deviates less than $1\%$ from the total variance of the density when consider up to $k \rightarrow \infty$. This is why all our S+R simulations have $k_{\rm{Ny}} / k_{\rm{h}} \gtrsim 4$ and why we only use N-body simulations for colder test cases as shown in Table \ref{tab:sims_cutoffs}. Further, with the mass resolution of $M_{\rm part} = 5.0 \cdot 10^{6} M_\odot / h$ the half-mode mass is very well resolved in our S+R simulations, by $M_{\rm{hm}} / M_{\rm{part}} \sim 10^3 - 10^4$ silent particles.

\section{Uncertainties of the Halo Mass Function} \label{sec:uncertaintyofMF}
\begin{figure*}
    \centering
    \includegraphics[width=\textwidth]{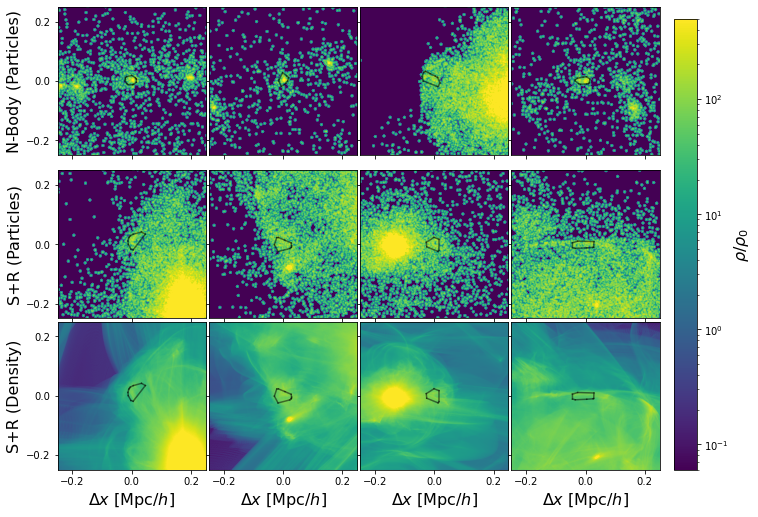}
    \caption{The density field around typical FoF groups. The different columns represent four randomly selected groups in the mass range $\SI{5e8}{\per\h\msol} < M < \SI{6e8}{\per\h\msol}$ ($\sim$ 100 particles). The top row displays the $1$ keV N-body simulation -- typical N-body FoF groups in this mass range correspond to artifical haloes. The central row shows groups in the $1$keV sheet simulations, but just displaying the $512^3$ silent particles used in the group-finding in order to compare fairly to the N-body case (also $512^3$). Typical FoF groups in this mass range  correspond to caustic structures in the outskirts of massive haloes. The bottom row visualises the same regions using the sheet-density field, showing that there are, in fact, no haloes at these locations. The colour bar has been clipped at 4e2. Black lines delineate the convex hulls of the FoF groups. The projection depth is $62.5 \rm{kpc}$ in all cases.}
    \label{fig:foffailures}
\end{figure*}

In this section, we will explore the shape of the halo mass function in our ``sheet + release'' (S+R) simulations. We show that the mass function in the strongly suppressed regime ($n_{\rm{X}} / n_{\rm{CDM}} < 5\%$)  depends sensitively on the definition of what is considered a halo, in particular, on the numerical details of the halo-finding algorithm. Large fractions of the objects detected by Friends-of-Friends halo finders and also of those considered bound by the \subfind{} algorithm \citep{springel2001} appear unbound when the tidal field is considered in the binding criterion. Further, many of these are unvirialized. We demonstrate these problems in Section \ref{sec:foffailures}; we introduce the new binding check from \citet{Stuecker_2021} that also considers the effects of the tidal field in Section \ref{sec:boostedpot}; and we investigate the virialization of haloes in Section \ref{sec:virialization}. Finally, in Section \ref{sec:uncertaintyMF} we consider the implied uncertainties in the abundance of very low-mass haloes, and we demonstrate that these primarily affect halo masses where the suppression exceeds a factor of 20 so that the abundance relative to CDM is well determined throughout the intermediate regime where it is of practical relevance. This suppression factor forms the basis of our quantitative analysis in Section \ref{sec:quantitativencdm}. In the current section we will focus on a single simulation, the 'normal' 1keV simulation.

\subsection{Unsuitability of traditional mass definitions} \label{sec:foffailures}
\citet{angulo2013} showed that, when applied to WDM simulations, the popular Friends-of-Friends (FoF) algorithm detects many small-scale over-densities which do not resemble conventional haloes. (This is a short-coming of the halo-finding algorithm and should not be confused with the issue of artificial haloes which reflects a short-coming of the simulation method.) We demonstrate the problem visually in Figure \ref{fig:foffailures} where we show random examples of FoF-groups in the mass range $\SI{5e8}{\per\h\msol} < M < \SI{6e8}{\per\h\msol}$ identified with a linking length equal to 20\%  of the mean interparticle separation. For the N-body simulations most (but not all) of the groups in this mass-range are artificial haloes spaced  regularly along filaments. However, for the S+R simulations the FoF-groups in this regime look very different. In the bottom panels of Figure \ref{fig:foffailures} we show the sheet-density estimate for these same regions and it becomes clear that the overdensities detected by the FoF-algorithm do not correspond to haloes, but rather to overdense shell or caustic regions in the outskirts of larger haloes.

The problem with the FoF algorithm is its assumption that all connected regions with over-densities above a certain threshold (e.g. $\rho/\rho_0 > 80$ for $l=0.2$; \citet{more_2011}) should correspond to virialized haloes. However, in the continuum limit $N \rightarrow \infty$ the density becomes arbitrarily high $\rho \rightarrow \infty$ near caustics \citep[e.g.][]{white2009, vogelsberger_2011} -- which are also abundant within and outside of haloes \citep[e.g.][]{feldbrugge_2018}. This effect can be neglected in most CDM simulations because the density field fragments on all scales and increasing the resolution decreases the density of the smooth component near caustics. N-body simulations with a power spectrum cut-off fragment artificially on small scales, reducing here also the likelihood of linking through caustics. However, from a theoretical perspective, all algorithms based on detecting haloes purely as overdensities are ill-defined and this becomes problematic in simulations with a small-scale cutoff, particularly if these are, like our S+R simulations, devoid of haloes formed by artificial   fragmentation.

\begin{figure}
    \centering
    \includegraphics[width=\columnwidth]{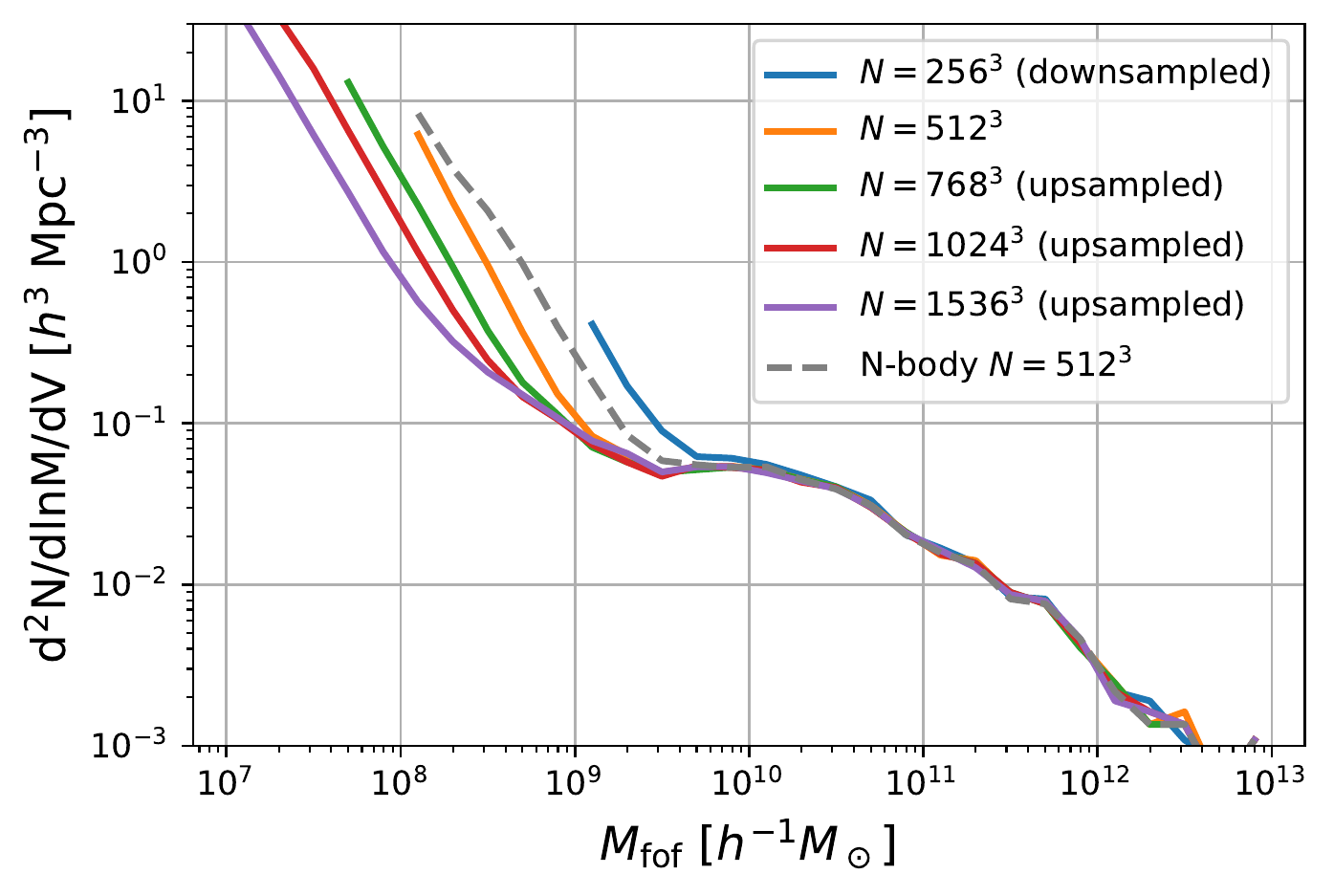}
    \caption{ Friends-of-Friends mass functions for different resampling levels. Each mass function is plotted down to objects with 20 resampled particles. The FoF algorithm identifies small-scale structures that are indeed physical features of the density field, e.g. high-density regions associated with  caustics (see Figure \ref{fig:foffailures}) but they are not at all similar to what we typically think of as haloes. The standard procedure of identifying haloes as compact, highly overdense regions does not work on small scale in WDM universes.}
    \label{fig:hmf_vs_resampling}
\end{figure}

We make a simple experiment to investigate this further. Using the sheet interpolation method on the $512^3$ silent particles of our S+R simulation, we resample its density field with an increasingly larger number of particles and then run a FoF group-finder on these new particle sets. The resulting mass functions can be seen in Figure \ref{fig:hmf_vs_resampling}. No matter how high the resolution, these seem to diverge at small masses. The low mass objects in this tail cannot be haloes of any kind, since the mass-resolution of the parent simulation is limited to $5 \cdot 10^6 M_\odot /h$. At much lower masses, the FoF mass function counts apparently disjoint peaks in overdensity which are not virialized quasi-equilibrium haloes. Such overdensities, like caustics, are not persistent material objects (particles move in and out of them over time) and  appear to connect smoothly to the cosmic web. An investigation of their statistics thus does not seem particularly useful.

A straight-forward idea for cleaning up the mass function is to adopt a structure-finder that also performs a binding check, the \subfind{} algorithm, for example. We do indeed find that many of the more curious looking FoF groups are unbound according to this binding check, which reduces the enhancement of the mass function at small masses. However, a visual inspection of the remaining seemingly bound objects shows that many still appear very unlike ``normal'' haloes. An example is shown in the first two panels of the second row of Figure \ref{fig:boosted_examples}. We find this to be a short-coming of the particular binding-criterion used by \subfind{} which incorrectly tags these objects as self-bound, as we will discuss next. Note that similar issues apply to all other structure-finders with a binding check that we know of.

To identify which particles are bound to a candidate halo or subhalo, \subfind{} considers the self-potential of the proposed members and discards those that have enough kinetic energy to escape from the group. This binding check neglects any effects due to tidal forces from surrounding material, even though these are expected to be substantial in many situations. Imagine that the self-gravity of a proposed group of particles creates a nearly spherical potential valley. Now, an external tidal field can deform this valley anisotropically -- bending the potential landscape downwards in one direction and upwards in another. If the tidal field is strong enough, it can even make the net potential curvature in one direction negative, thereby removing the minimum and the valley. An example of such a potential field is shown in the fourth panel of the second row of \ref{fig:boosted_examples}. Here, the equipotential lines open up along the filament. The tidal field is preventing the selected group of particles from further collapsing along the filament axis. We refer to \citet{Stuecker_2021} for an in depth discussion of this topic. We will discuss in the next section how to define such potential landscapes and how to use them to define a binding check that properly takes the tidal field into account.

\subsection{The Boosted Potential Binding Check} \label{sec:boostedpot}

\begin{figure*}
    \centering
    \includegraphics[width=\textwidth]{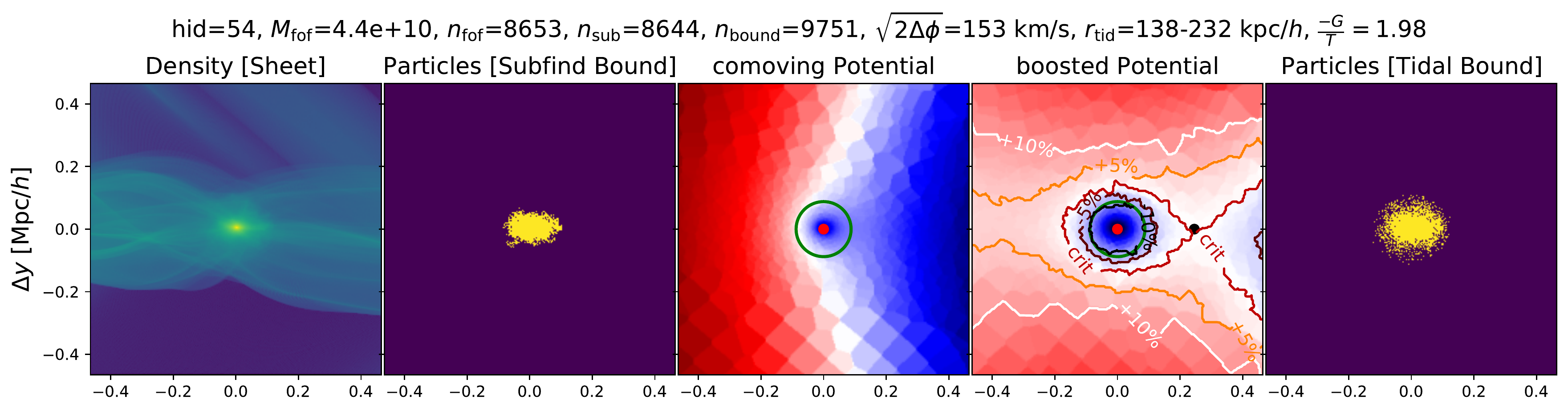}\\
    \vspace{-8pt}
    \includegraphics[width=\textwidth]{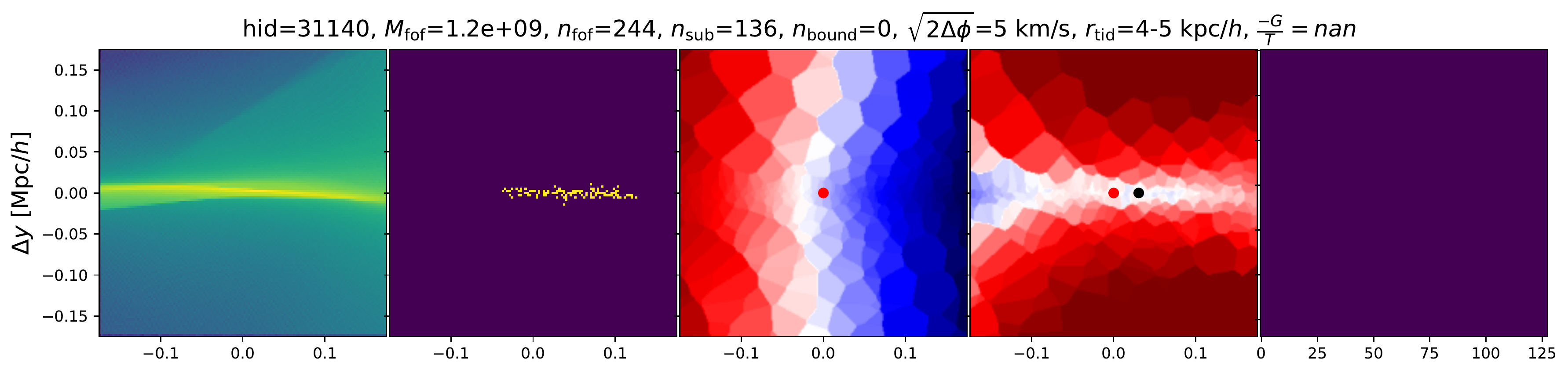}\\
    \vspace{-8pt}
    \includegraphics[width=\textwidth]{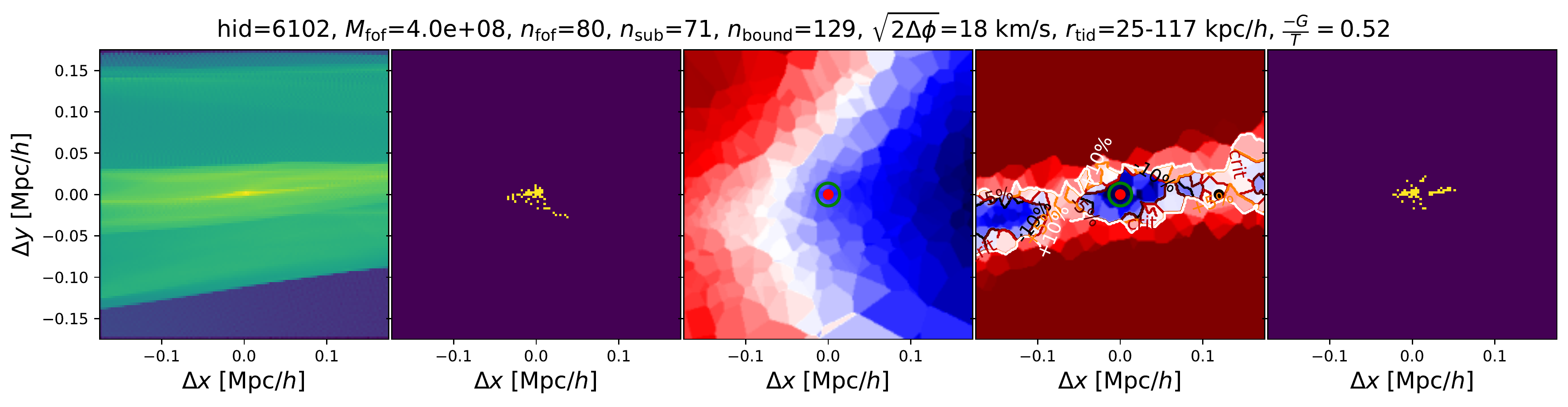}
    \caption{A few hand-picked examples to show how typical results from the \subfind{} binding criterion compare to those from our boosted-potential binding check. The first column shows the sheet-density around three structures identified as self-bound subhaloes by \subfind{}. The second and last columns show  particles that are considered to be members of these structures  by \subfind{} and by our boosted-potential algorithm, repectively. The third  shows the comoving potential $\phi$ while  the fourth shows a boosted potential obtained by subtracting a uniform gradient term (equation \ref{eqn:boostedpot}). \rvtext{The red point marks the \subfind{} position of the halo and the black point the  saddle-point of the boosted potential.} The examples, from top to bottom, are: a massive halo which is bound according to both criteria; 
    a caustic structure that is considered bound by \subfind{}, but is actually unbound due to the large external tidal field; 
    a structure which is bound according to both criteria, but which is not virialized, $-G/T = 0.52$.
    }
    \label{fig:boosted_examples}
\end{figure*}

In \citet{Stuecker_2021} we proposed a new binding check that accounts properly for the effects of tidal fields and is no more complex than traditional checks based on the self-potential of the proposed (sub)halo. We review the idea briefly here, but refer the reader to \citet{Stuecker_2021} for a more detailed discussion.


A natural quantity to consider for a gravitational binding check is the comoving potential inferred through the Poisson equation,
\begin{align}
    \Delta \phi &= 4 \pi G \left( \rho - \rho_0 \right), \label{eqn:poisson}
\end{align}
from the full density field of a cosmological simulation. Here $\rho_0$ is the mean cosmic matter density today, $\rho(\myvec{x},t)$ is the comoving density field at time $t$, and the Laplacian $\Delta$ is taken with respect to the comoving coordinate $\myvec{x}$. Cosmological definitions of gravitational potential can differ by factors of $a$; here we follow the convention of \citet{springel_2005}. The third column of Figure \ref{fig:boosted_examples}  visualises this potential in the vicinity of the three individual density structures shown in the first column. There is little apparent correspondence between the equipotentials of $\phi$ and the density distributions, but this largely reflects  the domination of $\phi$ by near-uniform gradients due to the large-scale density field. By the Principle of Equivalence, a  uniform acceleration does not affect the internal structure of a freely falling object, so that any potential of the form
\begin{align}
    \phi_{\rm{boost}}(\myvec{x}) &= \phi (\myvec{x})\; +\; \myvec{a_0} \cdot(\myvec{x} - \myvec{x}_0 ) \; -\; \phi_0 \label{eqn:boostedpot}
\end{align}
can provide a consistent description of the internal dynamics. \citet{Stuecker_2021} call a potential of the form given in eq.~(\ref{eqn:boostedpot}) a `boosted' potential and note that it is the natural choice for an object located at $\myvec{x}_0$ and freely falling with acceleration $\myvec{a}_0$. 

For each object previously identified by \subfind{}, we define a boosted potential through eq.~\eqref{eqn:boostedpot}, taking $\myvec{x}_0$ to be the centre of the subhalo (the point of highest density) and $\myvec{a}_0$ to be the acceleration
\begin{align}
    \myvec{a}_0 &= \langle \myvec{a} \rangle \label{eqn:boostedaverage}
\end{align}
averaged over the $0.8N$ closest particles, where $N$ is, for a  halo, the number of FoF group members, and, for a subhalo, the number of bound members according to \subfind{}. $\phi_0$ can be chosen arbitrarily. 

We visualise this boosted potential in the fourth column of Figure \ref{fig:boosted_examples}. A comparison with the third column clearly illustrates the advantages of the boosted potential. For well defined haloes, like the one in the first row, a halo boundary is naturally defined as the equipotential
passing through the saddle-point of the boosted potential (marked as a black point). This defines what would be commonly be called the `tidal-radius' of the halo, but can more usefully be thought of as bounding surface. Particles that have enough energy to cross this surface can exit the halo and never come back. More examples of boosted potential landscapes around haloes and subhaloes can be found in \citep{Stuecker_2021}.

The second row of Figure \ref{fig:boosted_examples} demonstrates that in regions where the tidal field is strong (like filaments), it is possible to have a large overdensity which appears bound according to its self-potential, yet does not actually correspond to a bound structure. By construction, the gradient of the boosted potential vanishes at $\myvec{x}_0$, but this will only correspond to a local minimum if all three eigenvalues of the tidal tensor (which is invariant to boosting) are positive. Here we use the `boosted potential binding check' as introduced by \citet{Stuecker_2021} to filter the halo-/subhalo-catalogue produced by FoF+\subfind{}. If suitably refined, the boosted potential could be used to define a structure finder based on the comoving potential alone, requiring neither the density field nor any arbitrary threshold, but we reserve this development for later work.

Our current version of the boosted potential binding check (BPBC) operates in three steps:
\begin{enumerate}[(i)]
    \item Determine the reference acceleration $\myvec{a}_0$ by averaging over a group of particles around a candidate center $\myvec{x}_0$.
    \item Find the critical contour of $\phi_{\rm{boost}}$ where the valley around $\myvec{x}_0$ merges with a deeper valley.
    \item Consider all particles inside the critical contour as candidate members of the (sub-)halo, but unbind all those that have enough kinetic energy to escape this surface.
\end{enumerate}
A more lengthy description with additional numerical details can be found in the appendix of \citet{Stuecker_2021}.

In the fourth column of Figure \ref{fig:boosted_examples} the critical contours are marked and we have normalized the colour scale so that the limiting potential is white, lower potentials are blue and higher ones are red. The fifth column of the same figure shows the bound particles according to the boosted potential binding check -- to be compared with the bound objects identified by \subfind{} and shown in the second column. In the top row the object is clearly more extended than what is suggested by \subfind{}. In the second row the \subfind{} object has no corresponding bound object in column five, reflecting the large effect of the tidal field.


We present halo mass functions inferred using different binding criteria in Figure \ref{fig:hmf_filtering}. Since we do not want to be sensitive to differences in mass definition, different lines in each panel are all inferred using the same definition, but in the various cases, we reject objects as unbound if less than $40\%$ of their reference mass is bound according to \subfind{} or to the boosted potential binding check.\footnote{The $40\%$ threshold is somewhat arbitrary. However, our results that depend on this threshold are mostly qualitative and we have checked that they are robust against variations of its value.} Differences between the mass functions in each panel are thus caused purely by  differences in the halo sets that pass the binding check. For the FoF mass function in the $1$keV WDM simulation (top panel), the filtering causes large differences, while for a CDM simulation (central panel) the differences are much smaller. In the bottom panel we apply the same filtering schemes to the WDM simulation, but now using a different mass definition $M_{200\rm{c}}$.  

Many small-mass FoF haloes  are considered unbound by \subfind{}, but the resulting mass function still has an upturn in abundance at the small-mass end. When the boosted potential binding check is applied, many of these objects are rejected as unbound, but a small upturn still remains, while at least the peak of the mass function becomes already clearly identifiable. As we will see in the next section, most of these objects are far from dynamical equilibrium. The bottom panel of Figure \ref{fig:hmf_filtering} shows that the number of low-mass objects is reduced in all cases if $M_{200c}$ is used instead of $M_{\rm fof}$, and in addition the differences between the cases become smaller.

\subsection{Virialization of WDM structures} \label{sec:virialization}

\begin{figure}
    \centering
    \includegraphics[width=\columnwidth]{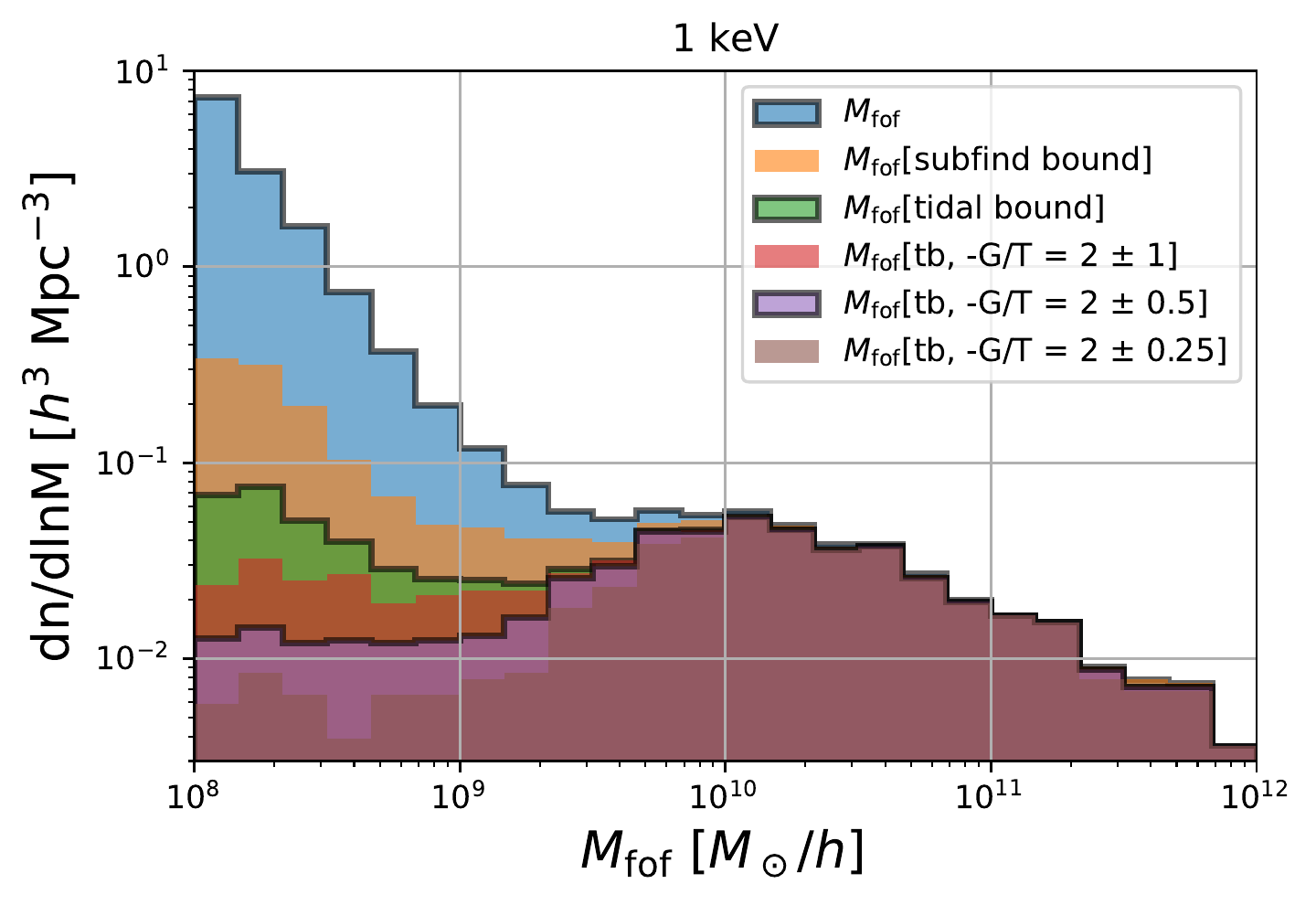}
    \includegraphics[width=\columnwidth]{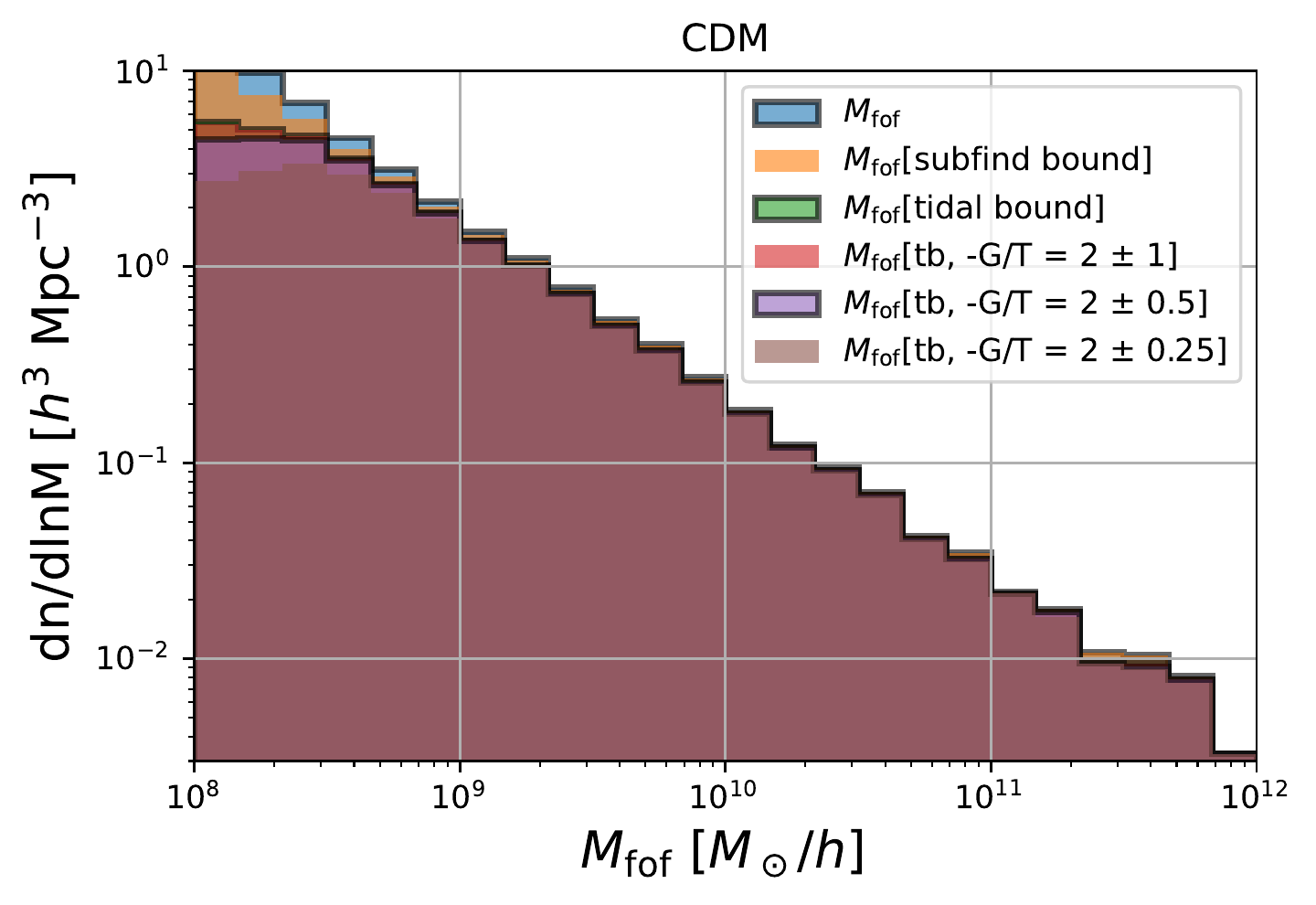}
    \includegraphics[width=\columnwidth]{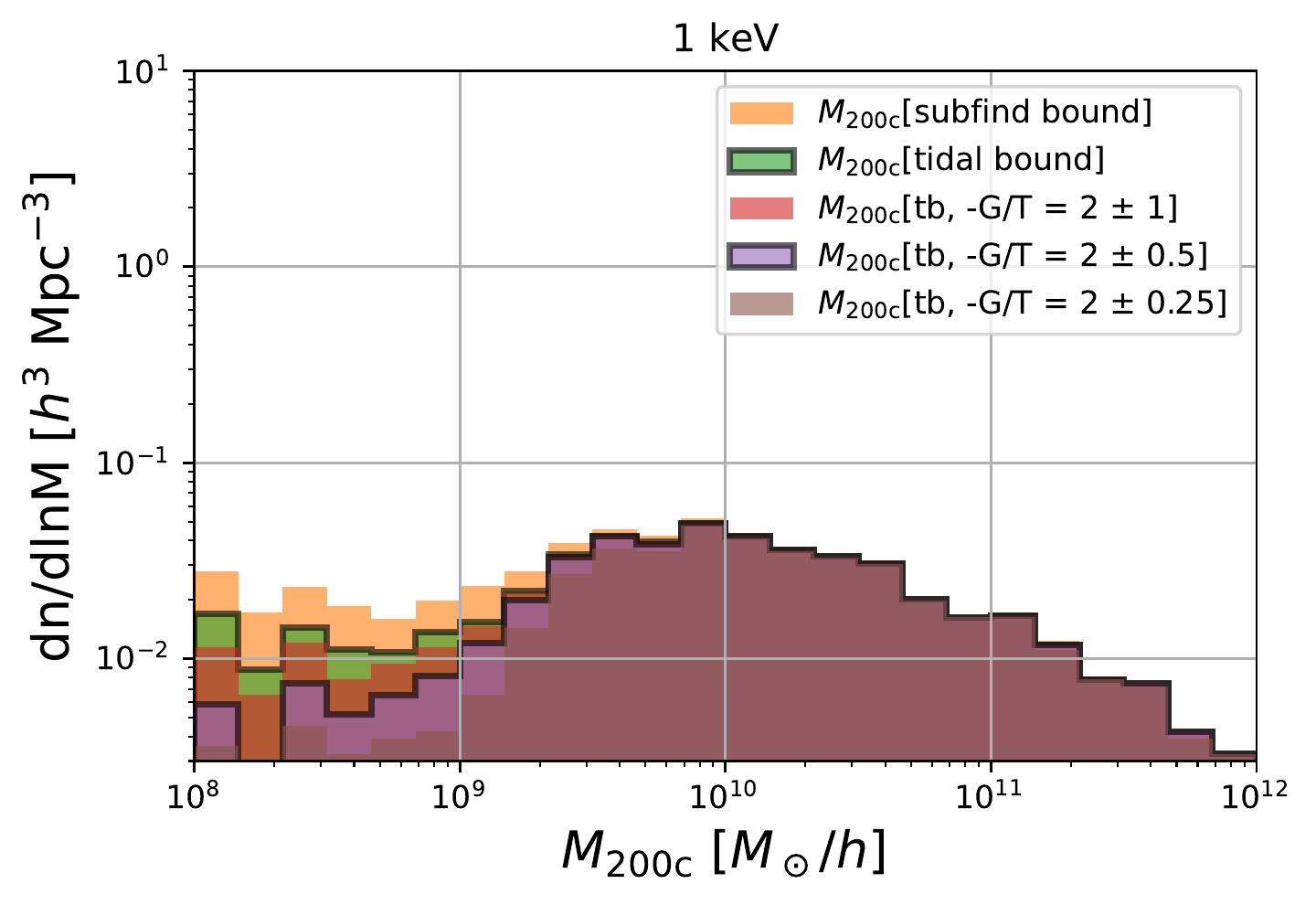}
    \caption{Top and central panels: FoF mass functions for the 1 keV WDM simulation and for a CDM simulation after filtering by various binding and virialization checks. There is considerable uncertainty arising from the definition of a halo. The CDM mass function is relatively robust to these differences, but in WDM the mass function is both highly uncertain and definition-dependent at low mass, even when no artificial haloes are present. Bottom panel: $M_{200,c}$ mass function in the WDM case filtered by the same set of criteria (for our simulations this is only defined for \subfind{} bound objects). This mass definition is considerably more robust, but still allows definition-dependent variations up to a factor of 5 or so.}
    \label{fig:hmf_filtering}
\end{figure}

\citet{angulo2013} argued that many of the small-mass objects in their simulations, are `protohaloes' -- i.e. objects that are in the process of forming and will grow rapidly in their subsequent evolution. Motivated by this, we test here whether the low-mass objects that pass the boosted potential binding check correspond to virialized haloes. 

Typically, the virialization of a halo is studied by comparing its self-potential binding energy with its internal kinetic energy. Here, however, we wish to avoid using the self-potential, and so use the virial theorem in its most basic form, defining the scalars
\begin{align}
    G &:= \left\langle \myvec{a} \cdot \left( \myvec{x} - \left\langle \myvec{x} \right\rangle \right) \right\rangle , \label{eqn:virialG}\\
    T &:= \frac{1}{2}\left\langle \left(\myvec{v} - \left\langle \myvec{v} \right\rangle \right)^2 \right\rangle,  \label{eqn:virialT}
\end{align}
where $G$ is the virial, and $T$ is the kinetic energy defined with respect to the centre of mass velocity $\langle \myvec{v} \rangle$. The scalar virial theorem then states that for a system in dynamic equilibrium,
\begin{align}
    - G\;/\;T &= 2. \label{eqn:virialtheorem}
\end{align}
For an isolated system, the virial $G$ can be converted to a potential energy, but we here test the virialization of objects directly through \eqref{eqn:virialtheorem}, which has the advantage that it  applies even when external forces are present. We compute the averages \eqref{eqn:virialG} and \eqref{eqn:virialT} over the particles that are bound according to the BPBC. We also tested the selection through the \subfind{} binding criterion, but it appears that generally the BPBC distinguishes more accurately which particles belong to the virialized halo, see \citet{Stuecker_2021} for more details.

We show an example of a structure that is bound according to all binding criteria, but not virialized ($-G/T \simeq 0.52$) in the last row of Figure \ref{fig:boosted_examples}. This object might indeed be interpreted as a protohalo that is in the process of collapsing, in agreement with the arguments in \cite{angulo2013}. 

Figure \ref{fig:hmf_filtering} includes BPBC filtered mass functions which are additionally filtered to require a virial ratio close to $2$ but with different allowed maximum offsets $\epsilon=$ 0.25, 0.5 and 1:
\begin{align}
    2 - \epsilon < -G/T < 2 + \epsilon
\end{align}
It is unclear what should be considered an appropriate threshold for virialization, but Figure \ref{fig:hmf_filtering} shows that many small-mass objects are far from equilibrium, so that the issue of virialization adds additional uncertainty when defining haloes at low mass.

\subsection{Uncertainties in Mass functions}
\label{sec:uncertaintyMF}

Figure \ref{fig:hmf_filtering} demonstrates that different halo definitions give very different results for the low-mass end of the halo mass function in WDM but agree reasonably well in the CDM case. Although this may, in part, be because most halo finders have been tested and tuned using CDM simulations, WDM objects in the strongly suppressed regime of the mass function are very different in nature from the clearly bound, nearly spherical quasi-equilibrium haloes that dominate the CDM mass function at all masses. As a result, in the WDM case the halo mass function at low mass depends strongly on the definition of a halo; Figure \ref{fig:hmf_filtering} should be understood as an indication of the strength of this definition-dependence. In Appendix B, we provide a similar figure for satellite subhaloes (Figure \ref{fig:shmf_filtering}) showing that a strong definition-dependence is also present in this case. We conclude that answering the question `{\it What is the shape of the halo mass function far below the half mode mass?}' requires first finding a clear and precise answer to the question `{\it What should we call a halo?}'

It may seem disappointing that although our simulations have eliminated the artificial haloes produced by N-body artifacts, they still do not provide an unambiguous measurement of the halo mass function at small masses. However, it is important to realise that this is a result of a change in the physical nature of structure below the cut-off scale.  In addition, we note that details of the mass function in this small-mass limit are quite irrelevant for most practical purposes. At the point where the objects found by  \subfind{} and the BPBC become qualitatively different, the mass function is already suppressed with respect to the CDM case by about a factor of 20. 

\begin{figure}
    \centering
    \includegraphics[width=\columnwidth]{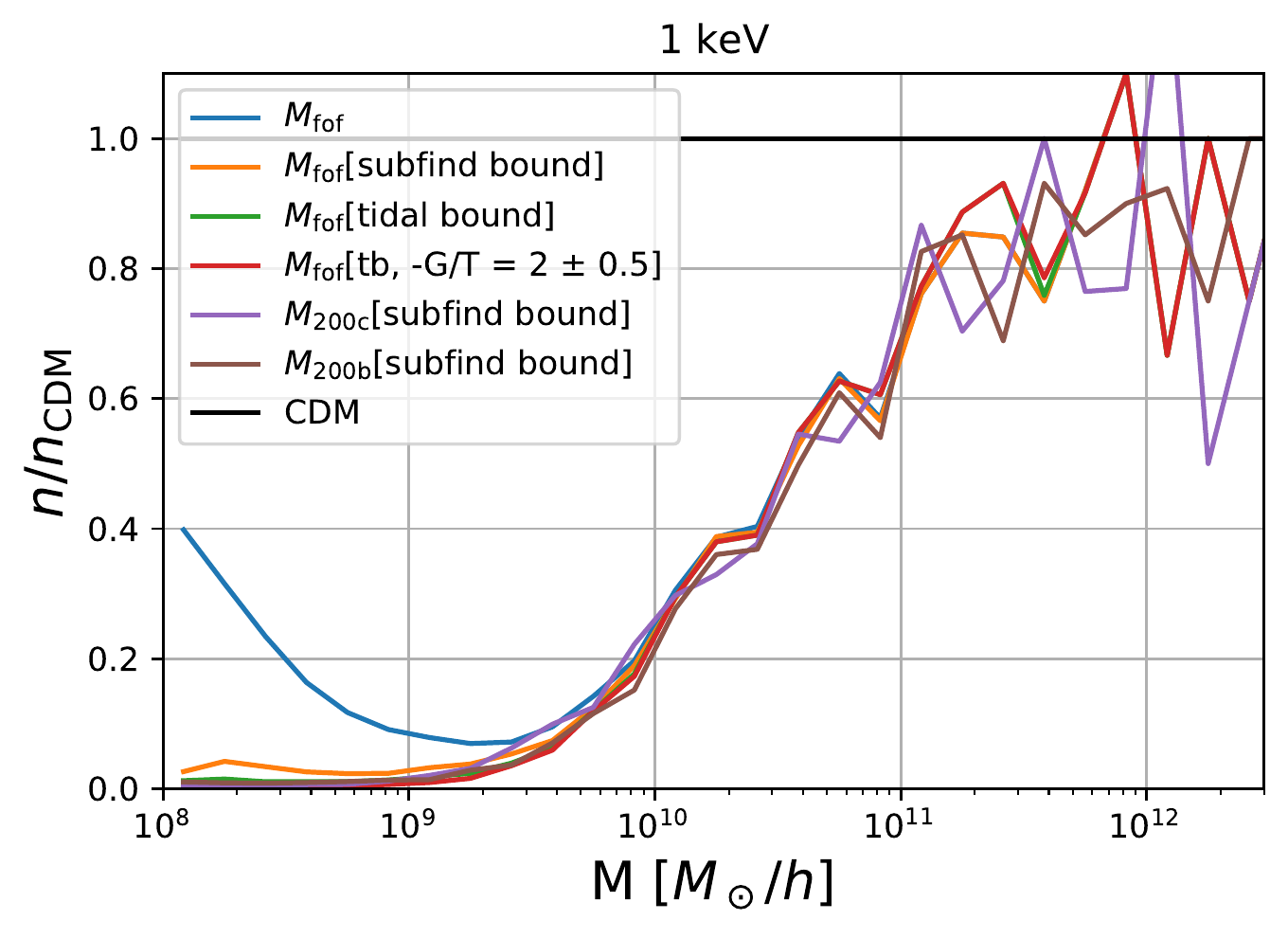}
    \caption{The suppression of the halo mass function in the 1 keV WDM case with respect to that in the CDM case for the various definitions of halo mass and the various filtering schemes presented in Figure \ref{fig:hmf_filtering}. When plotted on a linear scale, the relative suppression appears quite robust to the details of the halo definition as long as a binding check is imposed. Only the case of FoF masses with no binding check gives significantly discrepant results.  For the rest of this article we will investigate the relative suppression using $M_{200,c}$ (for haloes) and $M_{\rm{subfind}}$ (for subhaloes), since these are common choices in the literature and are easily reproducible for other researchers. See Figure \ref{fig:relativemassdefsubhalo} for a similar figure for subhaloes. }
    \label{fig:relativemassdef}
\end{figure}

This is illustrated in Figure \ref{fig:relativemassdef} which plots on a linear scale
the suppression of the mass function of our 1 keV WDM case relative to the CDM case for all the combinations of mass definition and binding check discussed above. All cases appear to give equivalent answers in this representation, except for the FoF mass definition with no binding check. Thus our new simulations give robust results for the primary purpose for which they are needed, namely to provide quantitative input to observational and experimental set-ups trying to distinguish WDM from CDM. A similar figure for satellite subhaloes can be found in Appendix B (Figure \ref{fig:relativemassdefsubhalo}) and is consistent with the same conclusion. We therefore concentrate on this relative suppression when discussing different dark matter models in Section \ref{sec:quantitativencdm}, where we also use $M_{200c}$ masses for haloes and $M_{\rm{sub}}$ masses for satellite subhaloes despite believing that the BPBC is preferable to \subfind{} for WDM simulations. This is because the two binding checks give similar results for the relative suppression and using the more common definition facilitates reproducibility.


\section{Quantifying the suppression relative to CDM for generic non-CDM models} \label{sec:quantitativencdm}
We have seen in the last section that the suppression of the halo and subhalo mass functions relative to CDM  can be reliably and robustly inferred in the intermediate suppression regime, $n_X(M) / n_{\rm{CDM}}(M) > 5\%$. In this section, we will introduce a generic expression for this relative suppression that it is suitable for use in future studies to predict the mass function for generic non-cold dark matter models. 

\subsection{Measurements} \label{sec:measurements}
\begin{figure*}
    \centering
    \includegraphics[width=\textwidth]{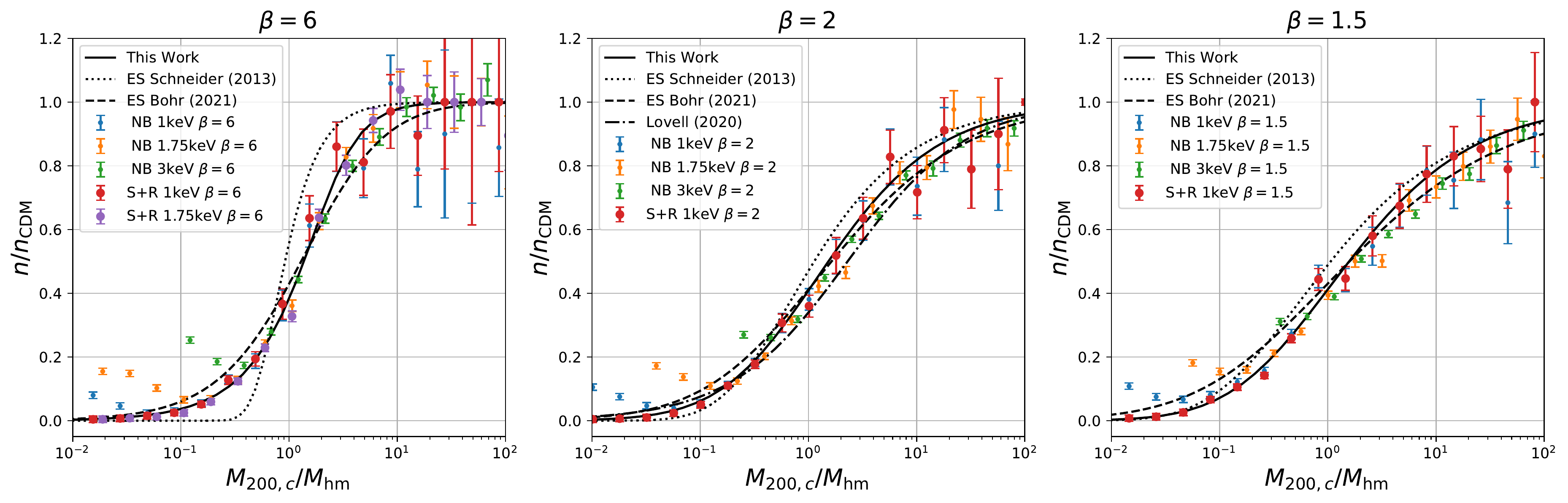} \\
    \includegraphics[width=\textwidth]{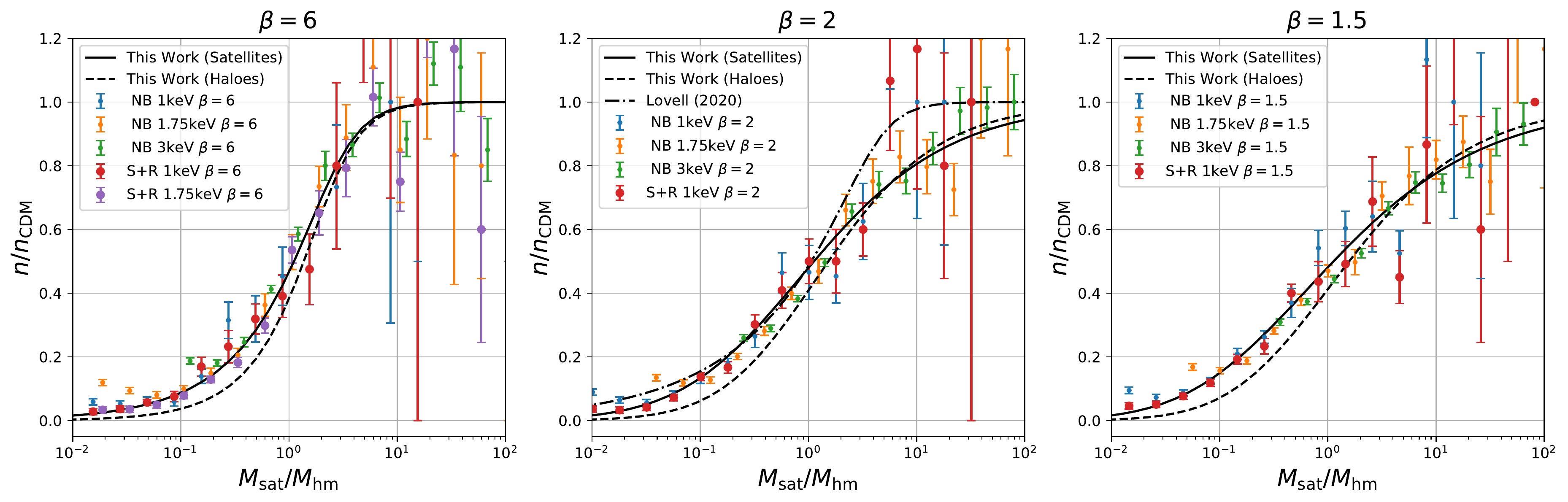}
    \caption{Our estimates of  halo (top) and satellite subhalo (bottom) abundance suppression ratios as a function of mass in units of the half-mode mass $\Mhm$ of equation \ref{eqn:halfmode}. From left to right, the panels correspond to different sharpness parameters $\beta$ of the power spectrum cutoff. Error bars are simulation-based jackknife estimates (see text). Solid black lines are fits using equation \ref{eqn:mscalesvsbeta}, while the other black lines in the top panel show previously published results. }
    \label{fig:relativemassfunctions}
\end{figure*}
In Figure \ref{fig:relativemassfunctions} we show mass function suppression ratios
\begin{align}
    f(M):=n_X(M)\,/\,n_{\rm CDM}(M)
\end{align}
 for haloes and for satellite subhaloes  for the various non-CDM models considered in this study. Masses ($M_{200c}$ for haloes, $M_{\rm{subfind}}$ for subhaloes) are given in units of the half-mode mass $\Mhm$ as defined in equation \eqref{eqn:halfmode}. Each suppression ratio is estimated in 20 equal logarithmic bins from $10^8$ to $10^{12} M_\odot / h$. Thus, each function is plotted down to (sub-)haloes containing 20 particles. The error-bars are determined using a jack-knife resampling where N=64 cubes with side-length $L/4$  are sequentially excluded so that 64 different instances of 63/64th of the volume are used to determine 64 values for the ratio $f_i$ in each mass bin. The errors $\sigma_f$ are then estimated using the standard jackknife estimator:
 \begin{align}
     \sigma_f^2 &= \frac{N-1}{N} \sum_{i=1}^{N} \left(f_i - \left\langle f \right\rangle \right)^2
 \end{align}
However, to avoid giving too much weight to the low-mass end, which has small statistical errors but substantial systematic ones (as discussed in the previous section), we impose a minimum errorbar of $0.01$, and we exclude all halos with fewer than 100 particles from our fits.

Figure \ref{fig:relativemassfunctions} shows the abundance suppression ratios to agree well for simulations with different cutoff scales, $\khm$, but similar sharpness parameter, $\beta$. However, they vary significantly as a function of $\beta$.
An enhancement at very low masses is evident in the N-body simulations, reflecting the presence of artificial haloes, but above the characteristic scale of these haloes, the suppression ratios estimated from N-body simulations agree well with those from `sheet + release' (S+R) simulations. Indeed the S+R simulations appear very similar to the artificial-fragmentation-free limit of the N-body simulations as estimated, for example, by \citet{Lovell2020}. Note that the agreement of simulations with different cutoff scales $\khm$ also shows that the measurements are relatively resolution and box-size independent, since these simulations have different particle numbers and abundance statistics at the same value of $M/M_{\rm{hm}}$.

\subsection{Fits and abundance suppression models} \label{sec:emulation}
\rvtext{Various approaches have been suggested in the literature to parameterize NCDM mass functions \citep[e.g.][]{schneider2012, lovell2014, hahn_2014, Corasaniti_2017, Lovell2020, bohr2021}, but many of these come with the drawback that they are only suitable to describe the suppression of the halo mass function, but not the subhalo mass function, which can be notably different \citep{Lovell2020}. We follow the suggestion of \citet{Lovell2020} to fit}
 the mass-dependent abundance suppression ratio, $n_X(M) / n_{\rm{CDM}}(M)$, with a function of the form
\begin{align}
    \frac{n_X(M)}{n_{\rm{CDM}}(M)} \simeq \left( 1 + \left(a \frac{\Mhm}{M} \right)^b  \right)^c.
\end{align}
\rvtext{where haloes and satellite subhaloes are each fitted individually with their own set of parameters.} 

In general, we find that functions of this form do indeed give a very good fit to our estimates of suppression ratios, but, as was also the case when fitting the power spectrum suppression using equation \eqref{eqn:transferabg}, the parameters of this function are highly degenerate -- i.e. one can easily find very different combinations $\{a,b,c\}$ which create very similar functions. This space is thus not well suited to finding regularities as a function of $\beta$. 

Because of this, we use the directly fitted parameters $\{a,b,c\}$ to infer the mass scales $M_{20\%}, M_{50\%}$ and $M_{80\%}$ where $n_X/n_{\rm{CDM}}$ equals $f=$ $20\%$, $50\%$, or $80\%$, respectively. This set of parameters has a one-to-one correspondence to $\{a,b,c\}$ through the system of equations,
\begin{subequations}
\begin{align}
    0.2 &= \left( 1 + \left(a \frac{\Mhm}{M_{20\%}} \right)^b  \right)^c, \label{eqn:abctomscale1}\\
    0.5 &= \left( 1 + \left(a \frac{\Mhm}{M_{50\%}} \right)^b  \right)^c,~~~{\rm and} \label{eqn:abctomscale2}\\
    0.8 &= \left( 1 + \left(a \frac{\Mhm}{M_{80\%}} \right)^b  \right)^c.  \label{eqn:abctomscale3}
\end{align}
\end{subequations}
We note that given $\{a,b,c\}$ it is easy to find $\{M_{20\%}, M_{50\%}, M_{80\%}\}$ through
\begin{align}
    \frac{M_{f}}{\Mhm} &= a \left( f^{1/c} - 1 \right)^{-1/b},
\end{align}
but the inverse mapping from $\{M_{20\%}, M_{50\%}, M_{80\%}\}$ to $\{a, b,c\}$ can only be inferred numerically. 

\begin{figure}
    \centering
    \includegraphics[width=\columnwidth]{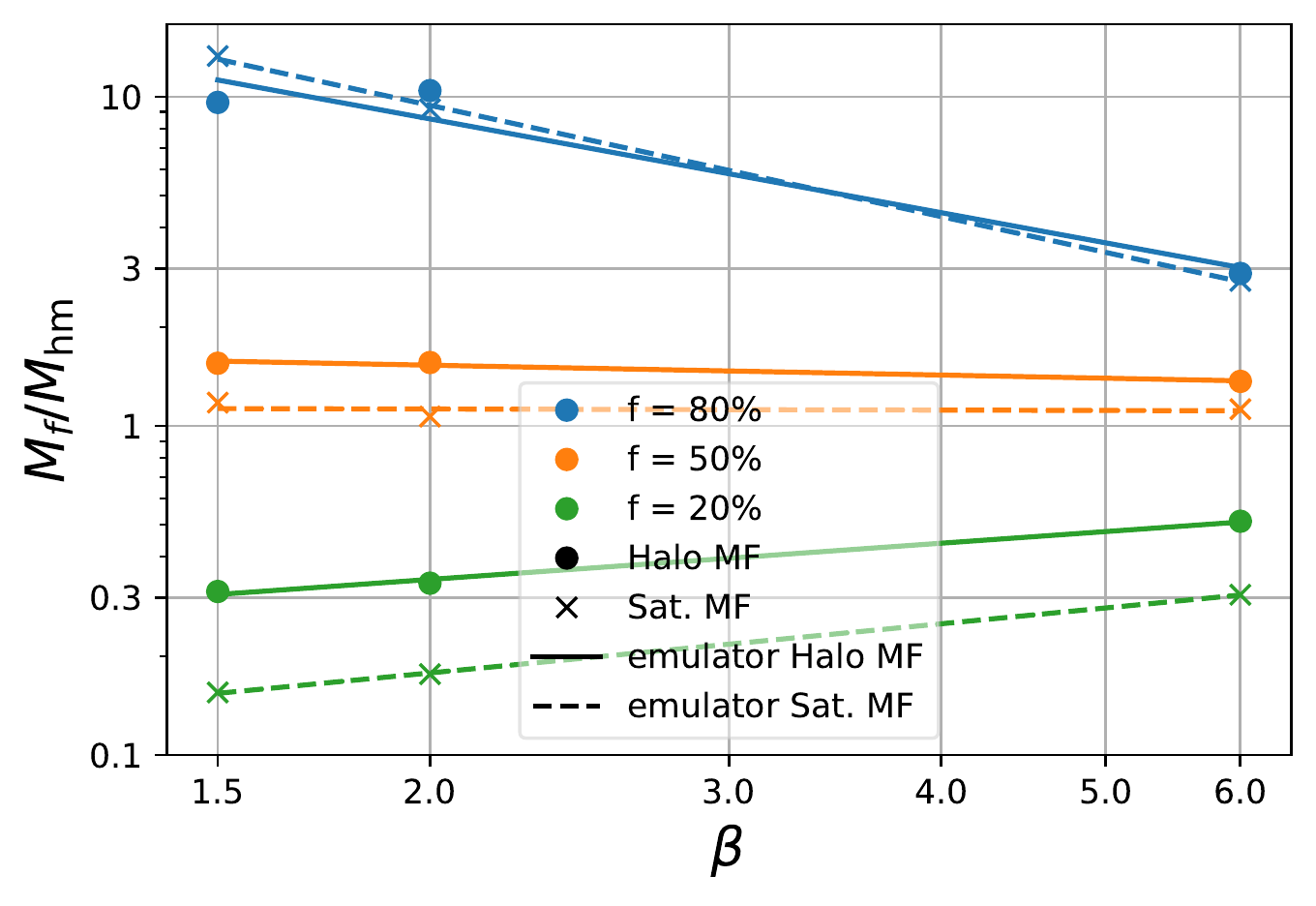} 
    \caption{Suppression mass-scales as a function of $\beta$. The dots and crosses indicate estimates based on the halo and satellite subhalo mass functions, respectively, while the lines indicate power-law fits as in equation \eqref{eqn:mscalesvsbeta}}
    \label{fig:emulator}
\end{figure}

We fit these functions in three different cases using only the S+R simulations:
\begin{itemize}
    \item $\beta = 1.5$ using the ``1kev-flat'' simulation,
    \item $\beta = 2$ using the ``1keV-normal'' simulation, and 
    \item $\beta = 6$  using both the ``1keV-sharp'' and the ``1.75keV-sharp'' simulations.
\end{itemize}
Figure \ref{fig:emulator}  shows the inferred  mass-scales as a function of $\beta$, the sharpness of the suppression of the initial power spectrum. The trend for each of the three mass scales seems simple enough to be approximated by a power law:
\begin{align}
    \frac{M_{f}}{\Mhm} = \mu_f \cdot \beta^{\nu_f} \label{eqn:mscalesvsbeta}
\end{align}
We fit such a power law to our measurements and list the resulting parameters, $\mu_f$ and $\nu_f$, in Table \ref{tab:mscales}. 

With these functions it is straightforward to estimate the mass function for most non-CDM cosmologies. First, one fits the high wave-number suppression of the initial power spectrum using the model of equation \eqref{eqn:transferabg} with $\gamma = 5$. Then one uses equation \eqref{eqn:mscalesvsbeta} with the values from Table \ref{tab:mscales} to infer the mass function parameters $\{M_{20\%}, M_{50\%}, M_{80\%}\}$ and maps these to $\{a,b,c\}$ by numerically inverting equations \eqref{eqn:abctomscale1} - \eqref{eqn:abctomscale3}. To simplify this procedure we provide a simple \textsc{python} script to do these steps in a code-repository along with some examples\footnote{\url{https://github.com/jstuecker/ncdm-mass-functions}}.
\begin{table}
    \centering
    \begin{tabular}{c|c|c|c|c}
      & f  & 20\% & 50\% & 80\%\\
     \hline
     Haloes & $\mu_f$ & 0.2651 & 1.638 & 16.51 \\
      & $\nu_f$ & 0.3656 & -0.0994 & -0.9466 \\
     \hline
     Satellites & $\mu_f$ & 0.1259 & 1.134 & 20.52 \\
      & $\nu_f$ & 0.4963 & -0.0110 & -1.1231 \\
    \end{tabular}
    \caption{Parameters  describing the suppression of the halo and satellite-subhalo mass functions a function of $\beta$ as in equation \eqref{eqn:mscalesvsbeta}.}
    \label{tab:mscales}
\end{table}

We show mass functions `emulated' in this way as black solid lines in Figure \ref{fig:relativemassfunctions}. By construction they provide a good fit to our S+R simulations. More significantly, they also fit well to colder N-body simulations, demonstrating that the mass function can indeed be scaled accurately in proportion to $\Mhm$.

\subsection{Interpretation and comparison with previous work}
It is interesting to see that the mass function cutoff behaves regularly as a function of the sharpness $\beta$ of the cutoff in the initial power spectrum, so that a two-parameter description is sufficient to represent the mass function shapes over the full range we have tested. We will test this further in the next subsection. It is also interesting to see that the satellite subhalo mass function (SMF) is very similar to the halo mass function (HMF) in the slightly suppressed regime $n/n_{\rm{CDM}} > 50\%$, but is more weakly suppressed at smaller mass. This is probably because the low-mass SMF is built in part from larger mass objects which have been tidally stripped, and so originated on more weakly suppressed scales.

A number of other studies have studied similar issues. \citet{Lovell2020} measured the halo and satellite mass functions from N-body simulations for the thermal relic case. The cutoff function they adopt (with $\beta\approx 2.2$, $\gamma \approx 4.5$) corresponds approximately to our $\beta=2$ case and is therefore shown only in the central panels of Figure \ref{fig:relativemassfunctions}. Our results agree very well with the HMF of \citet{Lovell2020}, but for the SMF we find good agreement only at low mass end, with a clear difference at weak suppressions where \citet{Lovell2020} find the SMF to be less suppressed than the HMF. It is hard to judge where this difference comes from, but we suggest that it may not be statistically significant, given the relatively small number of high-mass satellite subhaloes on which the SMF is based in each of our two studies. 

In addition, we compare with HMF predictions from two different excursion set (ES) approaches \citep[see e.g][]{press_1974, bond_1991, sheth_1999, sheth_2001, hahn_2014}. We first consider the ES approach presented by \citet{schneider2013} which uses a sharp k-space filter as a window function. We find that this approach gives good results for the $\beta = 2$ case. This makes sense, since this is a good representation of the suppression in the simulations that that the approach was fitted to. However, for the $\beta=6$ case this ES formalism produces much too sharp a cutoff in the HMF. This particular scheme appears to work only in the case for which it was explicitly tuned.

We next consider the ES formalism proposed by \citet{leo_2018} and applied in \citet{bohr2021}. This uses an ellipsoidal collapse model and a smooth window function of the form
\begin{align}
    W_R (k) &= \frac{1}{1 + \left( \frac{k R}{c_w} \right)^\nu}
\end{align}
where the parameters $c_w$, $\nu$ and a third parameter describing the ellipsoidal collapse are inferred through a fit to numerical data. When suggesting this window function, \citet{leo_2018}  argued that it describes non-cold dark matter models with a variety of cutoff shapes much better than the sharp k-space of \citet{schneider2013}. \citet{sameie_2019} and \citet{bohr2021} then applied it to simulations of various ETHOS models from \citet{vogelsberger_2016}. We adopt the parameter values of \citet{bohr2021} and compare the corresponding HMF suppression ratios to our own in the upper panels of Figure \ref{fig:relativemassfunctions}. We confirm that their model describes very well the variation of the HMF with the shape of the initial power spectrum cutoff. \citet{bohr2021} have also tuned the parameters of this ES formalism to reproduce the effect of the dark acoustic oscillations that appear in some of their ETHOS models. It is impressive that the same formalism  produces good results both for such dark acoustic oscillations and for variations in $\beta$. The formalism provides a prediction for the HMF only; the SMF must be inferred separately. Since our suppression ratio model is tested for both, we recommend using it for all cases where it is applicable, but in cases when this is not the case, for example, if the initial power spectrum cannot be well represented by our two parameter form, we recommend using the formalism of \citet{sameie_2019} with parameter values from \citet{bohr2021}. 

\subsection{Validation and high redshift}

\begin{figure*}
    \centering
    \includegraphics[width=\textwidth]{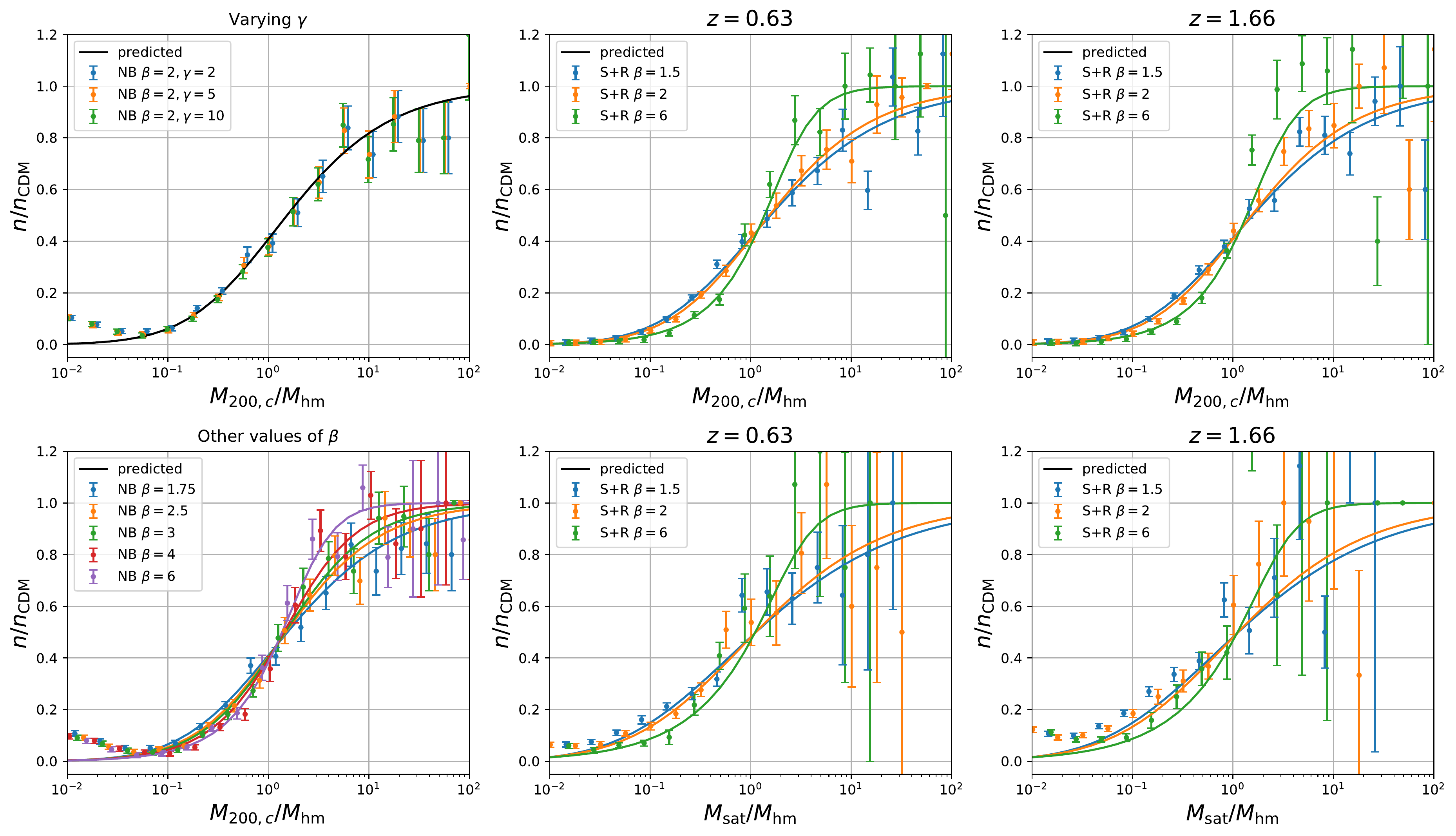}
    \caption{Validation plots for our suppression ratio model for a variety of deviations from the fiducial case. Top-left: varying $\gamma$ hardly changes the mass function, showing that the cutoff is well described by the proposed two-parameter model. Bottom left: the predicted functions for different values of $\beta$.  The model works well throughout the interval $1.5 \leq \beta \leq 6$.  Other panels: HMF and SMF suppression ratios at $z=0.63$ and 1.66. The model remains reasonably accurate for $z=0.66$, but clear discrepancies are visible at $1.66$. }
    \label{fig:validation}
\end{figure*}

So far we have shown that the emulation procedure discussed in Section \ref{sec:emulation} provides a good description of the mass functions of simulations with $\beta=1.5$, $\beta=2$ and $\beta=6$ and with substantial variations in half-mode scale. Figure \ref{fig:validation} explores how well the procedure works for other values of $\beta$ and for $z>0$. For these validation plots we use N-body simulations, since they are computationally cheaper and we have already shown them to agree well with the S+R scheme as long as results are used only above the scale of artificial fragmentation.

The top left panel of the figure demonstrates that if we keep $\beta$ fixed while varying the parameter $\gamma$ (which we have fixed to $\gamma=5$ so far), the mass functions barely change. This is because $\gamma$ has only a weak impact on the initial power-spectrum. It shows that our two-parameter description is sufficient for the problem at hand.

The lower left panel of Figure \ref{fig:validation} shows suppression ratios estimated from simulations with a range of different $\beta$ values and compares them with predictions from the emulation procedure described above. The predicted versus measured mass functions agree very well. 
Our emulation appears statistically compatible with all validation simulations for $\beta \in [1.5, 6]$. Given the validation plots from Figure \ref{fig:validation} and also the tests against N-body simulations with varying effective resolutions from Figure \ref{fig:emulator} we estimate that our model of the suppression factors is accurate at the level $\Delta f < 10\%$ in  this $\beta$ range.\footnote{Note that we \rvtext{advise} against applying the model outside of this range, e.g. for $\beta < 1.5$. In some preliminary test we find that for extremely flat cutoffs like $\beta = 1$ the low mass end gets underestimated. However, it is notoriously hard to make reliable simulations for such flat cut-off functions, since resolving the full cut-off requires a large dynamical range.}

Figure \ref{fig:validation} also explores whether the model for the HMF and SMF suppression ratios can be applied at higher redshift, in particular at  $z = 0.63$ and $z = 1.66$. We see that for both HMF and SMF the suppression ratios at $z = 0.63$ are well described by the model, but at by $z = 1.66$ some apparently significant discrepancies are visible in both cases.

\section{The Mass Concentration Relation} \label{sec:massconcentration}

Finally, in this section we will briefly consider the mass concentration relation (MCR) in our S+R simulations. The MCR is crucial for the inference of reliable constraints on the nature of dark matter. For example, \citet{gilman_2020} find that much of the constraining power on the WDM particle mass from flux-ratio anomalies in quadruply lensed quasars comes from its effect on the MCR, since concentrations are significantly reduced at halo masses well above the half-mode mass. There are two interesting questions that we would like to answer here: 
\begin{enumerate}[(a)]
\item \textit{Do S+R simulations and N-body simulations produce the same mass concentration relation?} So far all published results for the MCR are based on N-body simulations. However, it is possible that accretion of artificial haloes could alter the inner profiles of much more massive objects and so affect their concentrations. We can use our S+R simulations as an independent validation of previous results based on N-body simulations. 

\item \textit{Is the model of the MCR from \citet{ludlow2016} able to predict its variation as a function of the sharpness $\beta$ of the cutoff in the initial power spectrum?} The excursion set model proposed by \citet{ludlow2016} is widely used to model the MCR in universes with thermal-relic dark matter. However, if it is to be used to constrain other dark matter candidates, it is important to verify that it predicts the MCR reliably as a function of the shape of the small-scale cutoff in the initial power spectrum, parametrised in this paper by $\beta$.
\end{enumerate}
\begin{figure}
    \centering
    \includegraphics[width=\columnwidth]{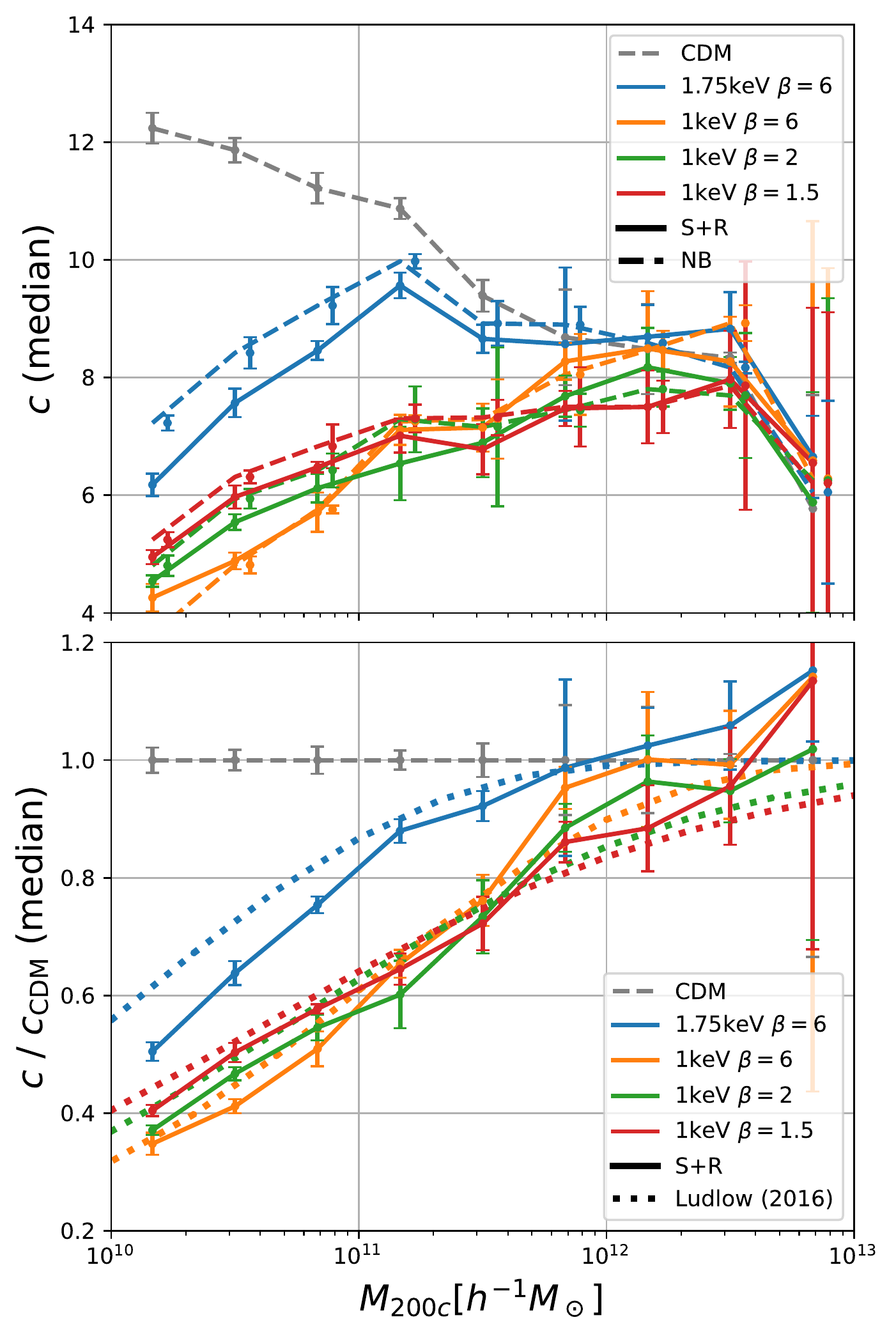} \\
    \caption{Top: A comparison of mass-concentration relations between the S+R and N-body simulations. Bottom: The ratio of the median concentrations of the NCDM cases to those of the CDM case in comparison to the model from \citet{ludlow2016}. Error bars are the statistical uncertainty of the median as determined through a jackknife technique. The error bars of the N-body simulations have been slightly displaced horizontally to avoid overlap.  Our S+R simulations independently confirm results from previous studies that were based on N-body simulations. }
    \label{fig:mconc}
\end{figure}
When estimating the MCR from our simulations, we only consider haloes with mass $M_{200c} > 10^{10} M_\odot/h$; these should be reliably resolved since they contain more than $2000$ particles. For each  halo we fit an NFW-profile \citep{nfw1996}
\begin{equation}
    \rho_{\rm NFW}(r) = \frac{\rho_cr_s}{r\left(1 + r/r_s\right)^2}
\end{equation}
in log-log-space in order to infer the concentration
\begin{align}
    c := \frac{r_s}{R_{200c}}
\end{align}
where $R_{200c}$ is the radius within which the mean density is 200 times the critical value. We divide haloes into nine logarithmic mass-bins from $10^{10}$ to $10^{13} M_\odot$ and compute the median concentration for each. Further, we determine the statistical uncertainty in these medians using the same jack-knife technique as in Section~\ref{sec:measurements}. 

The top panel of Figure~\ref{fig:mconc} compares the MCR's found from S+R and from N-body simulations. Overall it seems that the different schemes agree fairly well. For 1.75keV and $\beta = 6$, the MCR of the S+R simulation seems to be slightly lower than the N-body case at small mass. The internal scatter in the relation ($\Delta c \sim 2 $) is much larger than this difference, but the uncertainties we estimate for the median values do indicate a systematic difference which is not seen in any of the other cases. This is unlikely to reflect the effects of artificial haloes, which we would expect to be more pronounced in the warmer cases. However, it is still rather small, and a more sophisticated analysis of the convergence of halo profiles in the S+R simulations would be required to exclude any numerical effects. 

Overall, our S+R results are in good agreement with the results from N-body simulations. We find no  evidence that artificial haloes modify the MCR, thus validating previous results obtained with N-body schemes. We note that \citet{colombi_2021} has shown in a full phase-space analysis that N-body simulations give robust results when benchmarked against the \textsc{ColDICE} sheet-based simulation scheme \citep{sousbie2016}, at least when a large softening (exceeding one mean inter-particle separation) is used. While such a large softening would be generally desirable, it is hard to implement in practice, since a gigantic particle number would be required to resolve the centres of haloes. 
Our results suggest that N-body simulations with a small softening may give correct results for coarse-grained summary statistics like density profiles, mass-concentration relations and halo mass functions provided that the applicable range is carefully tested. Our S+R simulations establish that these statistics are not biased by artificial haloes, which are present in N-body simulations with a small softening but not in S+R simulations, but artifacts caused by discreteness effects \emph{inside} larger haloes cannot be excluded since these would be present in both types of simulation.

In the bottom panel of Figure \ref{fig:mconc} we plot the ratio of the median-concentrations of the NCDM and the CDM universes for our S+R simulations and compare them to the predictions of the excursion set model of \citet{ludlow2016}. Note that the high mass end might be affected by finite-box-size effects, yet the measured ratio between NCDM and CDM seems to be well described. For the 1keV cases we find excellent agreement. The model of \citet{ludlow2016} seems to capture the differences between the models as a function of $\beta$ very well, and it even reproduces the point where the concentrations cross at around $3 \cdot 10^{11} M_\odot$. This is quite remarkable, since \citet{ludlow2016} developed their model specifically for thermal relic universes ($\beta \approx 2$), and its use for other cases has never been tested (as far as we know). Only for the 1.75keV $\beta = 6$ case we again find that the MCR of the S+R simulations is below the prediction. As noted above, it is hard to judge the significance of this deviation, since this it is much smaller than the scatter around the MCR.

We conclude that our S+R simulations independently confirm the validity of the mass concentration relations that have been estimated. previously from N-body simulations. Further, we find that the model of \citet{ludlow2016} can be used to predict the MCR in universes with a cutoff that has different shape than in the fiducial thermal relic case, $\beta \approx 2$.

\section{Conclusions}

In this paper we have presented a first set of ``sheet + release'' (S+R) simulations of non-cold dark matter universes, specifically models in which the initial linear power spectrum is strongly suppressed on small scales. These simulations are free from the artificial fragments that appear in classical N-body simulations of such models and are able to follow the dynamics even in the centres of haloes where previous ``sheet'' simulation schemes break down. 

With these simulations we investigated the objects that form \rvtext{in} the strongly suppressed, low-mass tail of the halo mass function, finding them to be quite different from traditional haloes. The majority are elongated caustic overdensities that appear to be gravitationally unbound when the tidal field is considered in the binding check. In addition, we find objects which appear bound but not virialized, which may be protohaloes in an early stage of collapse, as suggested by \citet{angulo2013}. From this, we conclude that the halo mass function in the strongly suppressed regime $n_X(M)/ n_{\rm{CDM}}(M) < 5 \%$ cannot be inferred reliably, since it  depends too strongly on the precise definition of a halo. Any potential observable sensitive to the mass function in this regime would need to be inferred directly from the simulations, without specific assumptions about density structure or dynamical state, since for at least some observational probes the effect of the large number of caustic overdensities may  surpass that of the few traditional, quasi-equilibrum haloes at  low mass \citep[see e.g.][for the case of gravitational lensing]{richardson_2021}.

However, it is rather unlikely that the strongly suppressed tail of the mass function will ever be relevant for distinguishing between non-CDM candidates, since more easily measurable differences already arise in the regime of intermediate suppression where $n_X(M)/ n_{\rm{CDM}}(M) > 5 \%$. In the second part of this article we have focused on a quantitative description of halo and subhalo mass functions in this regime, finding that for most non-cold dark matter candidates the suppression of the primordial power spectrum can be adequately represented by just two parameters, namely $\khm$, which describes the spatial scale of the cutoff, and $\beta$, which describes its sharpness. We provide fitting formulae for the halo and subhalo mass functions as a function of these two parameters. These are applicable to a wide variety of non-CDM models, provided that the small-scale suppression of the initial power spectrum is well described by our two-parameter model.

Finally, we have found that the mass concentration relation for low-mass haloes is very similar in our S+R simulations and in standard N-body simulations, agreeing well with the model proposed by \citet{ludlow2016}. Further, we have tested this model for the first time for power spectrum cutoffs that are different in shape from that in the standard thermal relic case. We find that, in these cases also, it predicts the behaviour of the mass concentration relation astonishingly well for the range in sharpness parameter, $\beta$, that we have tested.

Overall, we find good agreement between our S+R simulations and classical N-body simulations for the summary statistics we studied.  We can confirm that the N-body scheme reliably predicts both the mass function and the mass concentration relation in non-cold dark matter scenarios when evaluated above the mass scale of artificial fragmentation \citep[see][]{wang2007} and above the most heavily suppressed regime of the mass function. 

\section*{Acknowledgements}

We thank Thomas Richardson for helping with evaluating and testing the simulations. J.S. thanks Marcos Pellejero-Ibañez and the BACCO group for interesting and motivating discussions. J.S. and R.A. acknowledge funding from the European Research Council (ERC) under the European Union's Horizon 2020 research and innovation program with grant agreement No. 716151 (BACCO). OH acknowledges funding from the European Research Council (ERC) under the European Union's Horizon 2020 research and innovation programme (grant agreement No. 679145, project `COSMO-SIMS'). The authors thankfully acknowledge the computer resources at MareNostrumIV and technical support provided by the Barcelona Supercomputing Center (RES-AECT-2019-3-0015).

\section*{Data Availability}

The data and algorithms underlying this article will be shared on reasonable request to the corresponding author. We provide some python scripts to facilitate the evaluation of the inferred mass functions in the \textsc{ncdm-mass-functions} repository under \url{https://github.com/jstuecker/ncdm-mass-functions}.

\bibliographystyle{mnras}
\bibliography{bibliography} 



\appendix
\section{Transfer functions fits}
\begin{figure*}
    \centering
    \includegraphics[width=\textwidth]{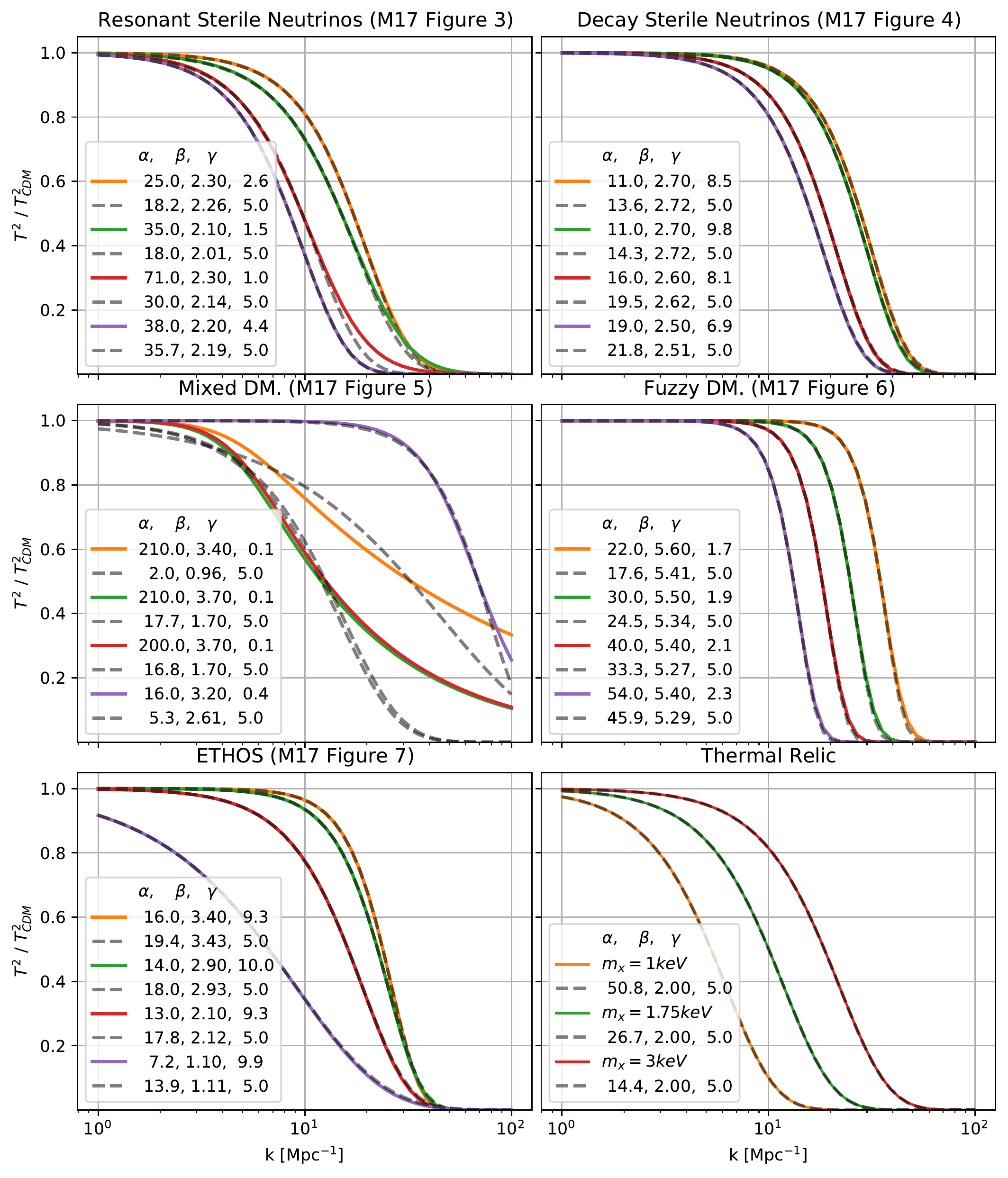}
    \caption{Transfer functions from \citet{murgia2017} when parameters are reduced from 3 $(\alpha, \beta, \gamma)$ to 2  $(\alpha, \beta)$. Solid lines correspond to three-parameter models and dashed lines to two-parameter fits with $\gamma=5$. The labels give $\alpha, \beta$ and $\gamma$ with $\alpha$ in units of $\rm{kpc}/h$. In all but the mixed dark matter case, the three-parameter model can be adequately reproduced varying just $\alpha$ and $\beta$.  For more detailed discussion of these dark matter models see \citet{murgia2017}.}
    \label{fig:murgia_fits}
\end{figure*}

In Figure \ref{fig:murgia_fits} we show a number of examples of dark matter models with a small-scale cutoff in the power spectrum that we have tried to represent with our two-parameter fitting function. These models are taken from \citet{murgia2017}, who show that their power spectum cutoffs are reasonably well approximated by the three-parameter ($\alpha,\beta$ and $\gamma$) model of equation \eqref{eqn:transferabg}. As explained in Section \ref{sec:transferfunctions}) and demonstrated explicitly in this figure, effectively equivalent behaviour is achieved by fixing $\gamma=5$ so that only two parameters remain. Only in the mixed dark matter case is $\gamma\neq5$ clearly required. 
For simplicity, the two- and three-parameter fits are matched in these plots by requiring agreement on the scales where $T^2$ is 0.5 and 0.85. With one exception, the models  considered by \citet{murgia2017} are almost equally well fit with two parameters as with three.

\section{Subhaloes} \label{app:subhaloes}
In this section, we show some complementary plots to the ones presented in Section \ref{sec:uncertaintyofMF}, but for the case of satellite subhaloes instead of haloes. 

\begin{figure}
    \centering
    \includegraphics[width=\columnwidth]{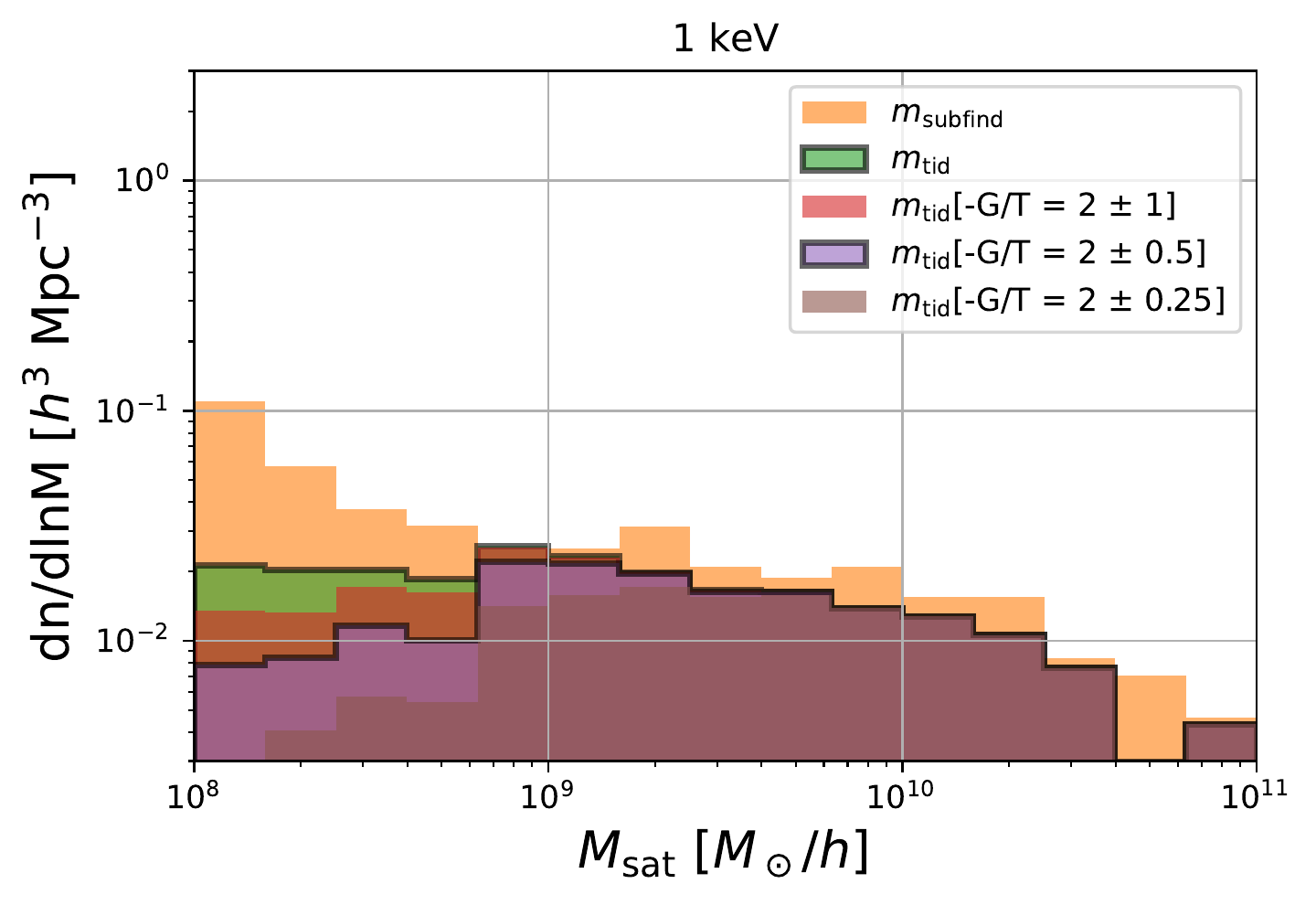}
    \includegraphics[width=\columnwidth]{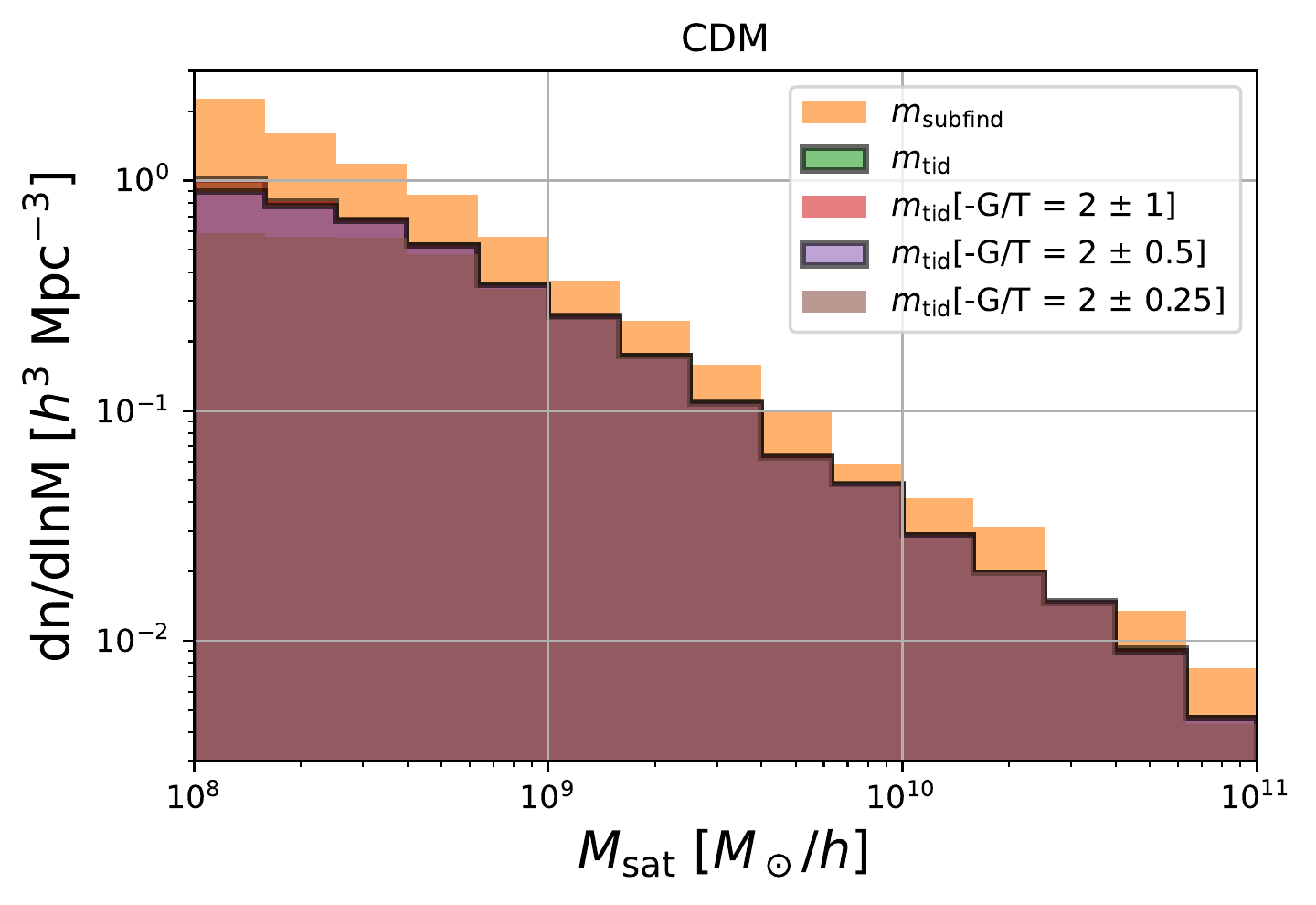}
    \caption{The satellite mass function for different filtering criteria / mass definitions in a WDM universe (top) and for CDM (bottom). In the CDM universe all of the mass definitions agree well, with the exception that there is an offset between the \subfind{} bound mass and the tidally bound mass (inferred through the boosted potential binding check). For the WDM case different filtering choices lead to very different results at the low-mass end. For example the unfiltered \subfind{} catalogue shows a steep rise at small masses whereas other filter choices show a flattening or even a decrease.}
    \label{fig:shmf_filtering}
\end{figure}

Figure \ref{fig:shmf_filtering} shows how the satellite subhalo mass function depends on mass definition and filtering criteria. It is directly analogous to the halo mass function plots of Figure \ref{fig:hmf_filtering}. Note that here we plot on the abcissa either the bound mass according to \subfind{}, or the bound mass according to the boosted potential binding check (BPBC). The counts of numbers in each bin are either unfiltered (in which case the \subfind{} mass is on the x-axis) or are filtered to require non-zero BPBC mass and various levels of virialisation (in which case the BPBC mass is on the x-axis). In the CDM case there is an almost constant offset between the \subfind{} and BPBC mass functions. For satellite subhaloes (which are typically subject to strong tidal fields) the BPBC typically finds fewer bound particles than \subfind{} -- on average the ratio is about $0.5$. Note that this is in contrast to haloes where the BPBC often finds larger masses than \subfind{}, since it does not depend on the FoF pre-selection.

For satellite subhaloes the lower end of the mass function depends strongly on the mass definition. The \subfind{} mass function turns up at small masses which is primarily due to unbound overdensities rather than actual subhaloes as shown by the difference between \subfind{}and BPBC masses.  When visually inspecting these objects, we found that most of them correspond to caustic-like structures in the outskirts of haloes. However, these effects are dominant in a mass range where the mass function is already suppressed by a factor of around 20 and are therefore probably irrelevant for quantitative studies which try to distinguish between different dark matter models.

\begin{figure}
    \centering
    \includegraphics[width=\columnwidth]{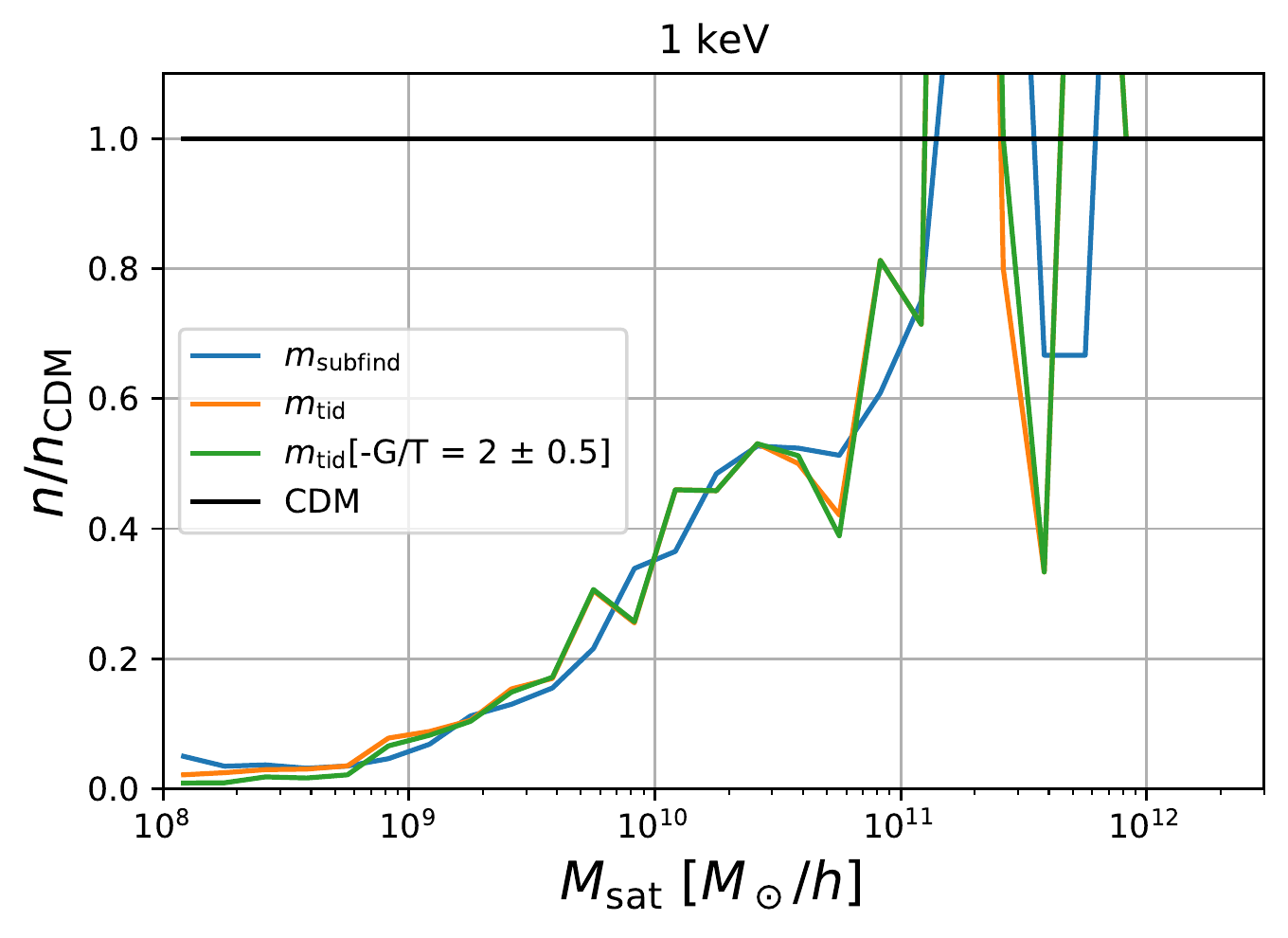}
    \caption{The relative suppression of the satellite subhalo mass function with respect to CDM for the various mass definitions presented in Section \ref{sec:uncertaintyofMF}. The relative suppression in the regime $n/n_{\rm{CDM}} > 5 \%$ is quite robust to the details of the mass definition. Compare with Figure  \ref{fig:relativemassdef} for haloes. }
    \label{fig:relativemassdefsubhalo}
\end{figure}

In Figure \ref{fig:relativemassdefsubhalo} we show the relative suppression of the satellite mass function for the different mass definitions and catalogue filterings, in direct analogy to Figure \ref{fig:relativemassdef}. As for the haloes, it becomes clear that the differences between the various cases appear small when considered on a linear scale and are substantial only where the SMF is suppressed by more than a factor of 20. Note, however, that  the \subfind{} mass function suppression does not seem to approach zero at small masses (corresponding to the upturn in Figure \ref{fig:shmf_filtering}). We consider this a failure of the mass definition to distinguish caustic overdensities from bound substructures.


\bsp	
\label{lastpage}
\end{document}